\pdfoutput=1
\documentclass[aps,prd,preprintnumbers,showpacs,twocolumn,groupedaddress,floatfix,nofootinbib]{revtex4-1}

\usepackage{latexsym} 
\usepackage{amssymb} 
\usepackage{amsmath} 
\usepackage{amsfonts}
\usepackage{bm}
\usepackage{color}
\usepackage{times} 
\usepackage{units}
\usepackage{hyperref}
\usepackage[utf8x]{inputenc}
\usepackage{graphicx} 
\usepackage[squaren]{SIunits}

\newcommand{\Fref}[1]{Fig.~\ref{#1}} 
\newcommand{\Sref}[1]{Sec.~\ref{#1}} 
\newcommand{\Tref}[1]{Table~\ref{#1}} 
\newcommand{\msec}{\usk\milli\second} 
\newcommand{\parsec}{\mathrm{pc}}

\begin{document}

\title{Structure of stable binary neutron star merger remnants: Role of initial spin}

\author{W. Kastaun}
\affiliation{Max Planck Institute for Gravitational Physics (Albert Einstein Institute), \\
Callinstr. 38, 30167 Hannover, Germany}
\affiliation{Leibniz Universität Hannover, Institute for Gravitational Physics, \\
Callinstr. 38, 30167 Hannover, Germany}
\affiliation{Physics Department, University of Trento, 
via Sommarive 14, I-38123 Trento, Italy}
\affiliation{INFN-TIFPA, Trento Institute for Fundamental Physics and Applications, \\
via Sommarive 14, I-38123 Trento, Italy}

\author{R. Ciolfi}

\affiliation{INAF, Osservatorio Astronomico di Padova, 
Vicolo dell’Osservatorio 5, I-35122 Padova, Italy}

\affiliation{INFN-TIFPA, Trento Institute for Fundamental Physics and Applications, \\
via Sommarive 14, I-38123 Trento, Italy}

\affiliation{Physics Department, University of Trento, 
via Sommarive 14, I-38123 Trento, Italy}

\author{A. Endrizzi, B. Giacomazzo}

\affiliation{Physics Department, University of Trento, 
via Sommarive 14, I-38123 Trento, Italy}

\affiliation{INFN-TIFPA, Trento Institute for Fundamental Physics and Applications, \\
via Sommarive 14, I-38123 Trento, Italy}

\begin{abstract}
We present general relativistic numerical simulations of binary neutron 
star (BNS) mergers with different initial spin configurations. We focus 
on models with stars of mass $1.4 M_\odot$ each, which employ the equation of 
state (EOS) by Shen, Horowitz, and Teige, and which result in stable NSs as
merger remnants. For comparison, we consider two irrotational equal mass 
($M=1.35\usk M_\odot$) and 
unequal mass ($M=1.29,1.42\usk M_\odot$) 
BNS models using the APR4 EOS, which result in a supramassive 
merger remnant. We present visualizations of the fluid flow and temperature 
distribution and find a strong impact of the spin on vortex structure
and nonaxisymmetric deformation. We compute the radial mass distribution
and the rotation profile in the equatorial plane using recently developed 
measures independent of spatial gauge, revealing slowly rotating cores that 
can be well approximated by the cores of spherical stars. We also study the 
influence of the spin on the inspiral phase and the gravitational wave (GW) 
signal. Using a newly developed analysis method, we further show that gravitational 
waveforms from BNS mergers can exhibit one or more phase jumps after merger, 
which occur together with minima of the strain amplitude. We provide a natural 
explanation in terms of the remnant's quadrupole moment, and show that cancellation
effects due to phase jumps can have a strong impact on the GW power spectrum.
Finally, we discuss the impact of the spin on the amount of ejected matter.
\end{abstract}


\maketitle

\section{Introduction}
\label{sec:intro}

After the recent detections of gravitational waves (GWs) from binary black hole (BBH) merger 
events by the LIGO detectors \cite{LIGO:BBHGW:2016,LIGO_GW151226} finally opened the 
era of GW astronomy, there is hope to detect also GWs from binary neutron star (BNS) 
mergers in the near future. The main uncertainty is the frequency of such events inside 
the detection horizon, see \cite{Abadie2010,LIGO:2016:832L21} for observational upper limits
and a review of current estimates.

Modeling the waveforms of BNS mergers is more complex than for the BBH case. During the 
late inspiral, there can be tidal effects that depend on the NS deformability (see 
e.g. \cite{Hinderer:2016:181101, Bernuzzi:2015:161103, Hotokezaka:2015:064060}) which in 
turn depends on the equation of state (EOS) for cold matter at supranuclear densities. 
This might help to infer information about the latter.
However, NSs can also have spins. Although the spin cannot be as large as for BHs, recent studies
\cite{Kastaun:2013:21501,Kastaun:2015:064027,Tsatsin:2013:64060,Dietrich:2016:044045} have 
found a significant impact on the late inspiral, comparable to tidal effects.
It is however unclear how rapidly NSs in binaries close to merger typically rotate. The 
component J0737−3039A of the famous double pulsar system \cite{Burgay:2003:426} reaches 
$44\usk\hertz$. Using the current spindown and assuming a constant braking index, it 
was estimated in \cite{Tichy:2011:24041,Dietrich:2015:124007} that the spin
at merger will still be ${\approx}37\usk\hertz$.
However, too few BNS systems have been observed to exclude larger spins.
The possibility of initial NS spin therefore needs to be considered for the interpretation 
of BNS GW signals. 

We note that there are only a few studies on the spin since suitable methods of computing 
initial data have been developed only recently. 
In \cite{Tsatsin:2013:64060,East:2016:024011,Paschalidis:2015:12} solutions of isolated 
rotating stars were superposed. In \cite{Tsatsin:2013:64060}, the constraint violating data 
was evolved, while in \cite{East:2016:024011,Paschalidis:2015:12} the constraint equations were solved again after 
the superposition. In \cite{Clark:2016:94}, the Bowen-York formalism for BBHs was extended 
to BNS. In \cite{Kastaun:2013:21501, Kastaun:2015:064027}, a rotational 
velocity fields was added manually to irrotational binary models and the constraint violating initial
data was evolved with a constraint-damping formalism.
In
\cite{Dietrich:2016:044045, Tacik:2015:124012, Dietrich:2015:124007, Bernuzzi:2014:104021,Dietrich:2017:1706.02969} fully 
consistent initial data was used, computed with the method developed 
in \cite{Tichy:2011:24041,Tichy:2012:064024}. 
Most studies considered aligned spins, with the exception of 
\cite{Tacik:2015:124012,Dietrich:2015:124007} who present precessing systems as well.
Further, \cite{Kastaun:2015:064027} is the only study of spinning stars that also uses 
three-parametric tabulated nuclear physics EOSs including thermal and composition effects.

In contrast to the BBH case, the waveforms from BNS mergers do not necessarily terminate at 
merger. If the total mass is below a certain threshold 
\cite{Hotokezaka:2011:124008,Bauswein:2013:131101}, the merger results in a strongly 
deformed remnant that emits GWs with strain amplitude comparable to the last orbit, but 
higher frequencies. This postmerger signal carries a wealth of information.
The most basic quantity provided by future detections 
of a postmerger signal
will be a lower limit on the 
lifetime of the remnant (lower limit because the signal amplitude can simply fall below 
detection threshold without collapse to a BH). 
The lifetime of the remnant depends on its mass. Roughly, one classifies
merger remnants as hypermassive if the total mass of the initial binary is above the 
maximum mass supported by uniform rotation (but below the threshold for prompt collapse 
during merger), and supramassive if the mass is above the maximum mass for a nonrotating
NS (Tolman, Oppenheimer, Volkoff (TOV) solution). If it is even lighter, the remnant is a stable NS. All the above thresholds
depend strongly on the EOS. 

An observational constraint of the remnant lifetime together with the total mass should 
also constrain the EOS. The above classification is however only a rough indication, since 
the actual remnant lifetime depends on many factors. The first factor is the mass of the  
disk that forms around the remnant, and its accretion time scale. Another important 
factor is the rotation profile of the remnant. A common notion of hypermassive stars is 
that differential rotation with a rapidly rotating core (as described in \cite{Baumgarte:2000:29}) 
is preventing immediate collapse, and that collapse occurs when the 
differential rotation inside the remnant falls below a critical level. Similarly,
supramassive remnants are thought to be stabilized by uniform rotation of the core, and to
collapse once sufficient angular momentum has been carried away by GW and/or magnetic braking.
A mounting number of studies 
\cite{Kastaun:2015:064027, Endrizzi:2016:164001, Kastaun:2016, Hanauske:2016}
find however that the cores of hypermassive and supramassive merger 
remnants rotate more slowly than the outer layers.
In \cite{Kastaun:2015:064027}, it was shown that the outer layers can approach Kepler velocity,
hence reducing the pressure onto the core and avoiding collapse. In this scenario, the 
lifetime depends on the angular momentum balance of those outer layers. Only one study 
\cite{Kastaun:2015:064027} has addressed the impact of initial NS spin on the rotation profile 
so far, for the case of two hypermassive models. In this work, we study the same for
a model in the stable remnant mass range.

The second feature likely to be observed by GW astronomy is the dominant frequency of the 
early postmerger stage ($<20\usk\milli\second$), where the signal is still strong.
This frequency is determined by the remnant properties, which are mainly given by mass and EOS,
but might also be influenced by mass ratio and spins. So far, there is no robust model for the
early remnant, which necessitates nonlinear numerical simulations.
In a recent publication \cite{Kastaun:2016}, we studied the fluid flow of a remnant in the stable 
mass range in detail
and found that directly after merger the fluid flow cannot be described as simple differential 
rotation. In a suitable rotating frame, we found a strong, but slowly varying density deformation,
with a fluid flow that contained two large vortices which rotated against each other before merging 
into one vortex. Even at this point the flow was not simple differential rotation; we observed
large secondary vortices, phase locked with the main density deformation and also related to hot 
spots. We will repeat the analysis of the irrotational model shown in \cite{Kastaun:2016} 
for cases with different aligned NS spin configurations.

The interaction between vortices, hot spots, and the nonaxisymmetric deformation of the remnant 
is not well understood yet. One open question is to what degree the changes of the remnant deformation 
are caused by the dynamics of secondary vortices, as opposed to angular momentum loss.
In any case, those findings suggest that linear perturbation theory is not a suitable tool
to describe such systems, in particular the main GW frequency. A robust model of the remnant
is required in order to constrain the EOS from observations of postmerger GW frequency.
There are phenomenological studies 
\cite{Bauswein:2012:11101,Bauswein:2012:63001,Hotokezaka:2013:44026,Bauswein:2014:23002,Takami2015} 
linking the remnant oscillation frequencies to properties of the initial stars or to the tidal 
coupling constant of the binary \cite{Bernuzzi:2015:091101}, but those studies do
not consider the initial NS spin.

The analysis of BNS waveforms is complicated by rapidly changing frequencies of the main signal.
The phase velocity often shows pronounced peaks during and/or shortly after merger, which often correlate
with sudden minima in the amplitude. In a previous work \cite{Kastaun:2015:064027}, we interpreted the 
frequency maximum during merger as a consequence of the maximum compression. In this work, we will discuss 
another effect that naturally leads to sudden frequency minima and maxima without requiring sudden physical 
changes, namely near-zero crossings of the remnant quadrupole moment in a rotating frame.
For this, we developed a new method to decompose the complex-valued GW strain obtained in numerical 
simulations.
In some cases, the strain amplitude also shows a more complex evolution than simple damping. 
In particular, one can often observe temporary minima. 
One might attribute those to unstable growing modes (this was suggested
e.g. in \cite{Feo:2017} as a possible explanation for temporary minima encountered in their
simulation results). Without identification of the mode, unstable mode
growth remains speculative. It is also not the only explanation: we will
discuss rearrangements of the fluid flow as another possible explanation
for similar minima encountered in our simulations.

This paper is organized as follows. In \Sref{sec:models}, we introduce our models and discuss potential shortcomings
as well as a method to quantify the spin of the stars in general relativity (GR). In \Sref{sec:code}, we describe the evolution code and
diagnostic quantities used. Our new GW analysis methods are described in Secs.~\ref{sec:gw_extr} and \ref{sec:jump_det}. 
The numerical results are presented in \Sref{sec:results}. 
Unless noted otherwise, we use units $G=c=M_\odot=1$.

\section{Setup}
\label{sec:numerics}

\subsection{Models}
\label{sec:models}
In this work, we mainly investigate an equal mass binary with the Shen-Horowitz-Teige (SHT)
EOS \cite{Shen:2010:15806,Shen:2011:35802}
and four different configurations for the initial spin of the NS: irrotational (IRR),
both aligned (UU), both antialigned (DD), and aligned-antialigned (UD) with respect to 
the orbital angular momentum.
The irrotational model has already been investigated in \cite{Kastaun:2016}.
The parameters of the models are listed in \Tref{tab:init_param}. 
Due to the large maximum baryonic mass ($3.33\usk M_\odot$) of TOV stars supported by the 
SHT EOS, the merger remnant 
will be a stable NS. This is a corner case, but might well occur in nature; at least it is
not ruled out by any observation yet.

In order to assess the generality of some results not related to spin, we also
consider two irrotational models with a piecewise polytropic approximation 
\cite{Read:2009:124032} of the APR4 EOS \cite{Akmal:1998:1804}. During evolution, 
a $\Gamma$-law thermal part with $\Gamma_\mathrm{th}=1.8$ is added. 
Both models are in the supramassive mass range for the APR4 EOS.
The unequal mass model is the one studied in \cite{Endrizzi:2016:164001}, and the 
equal-mass model is presented in detail in a different publication \cite{Ciolfi:2017:063016}; 
here we analyze the same simulation data in comparison to our new results.

All models are produced using the \textsc{LORENE} 
code \cite{Gourgoulhon:2001:64029,LoreneWWW}.
In order to add initial spin, we follow the receipt in 
\cite{Kastaun:2013:21501,Kastaun:2015:064027}. In short, we add a rotational 
velocity field to the irrotational initial data. The velocity field 
is obtained by scaling the residual velocity field in the coorbiting frame
by a constant factor.

This method of adding spin has mainly three potential sources of error. 
First, we violate the constraint equations of GR. In particular, the metric
is completely unchanged on the initial time slice, thus missing spin related 
effects such as frame dragging.
Second, the rotational velocity field added is not \textit{per se} guaranteed
to conserve the linear momentum of the stars. Such an error would manifest 
itself as differences of the orbital eccentricity between the different spin
configurations. 
Finally, since we do not change the mass density, hydrostatic equilibrium is not 
respected, which will lead to oscillations of the NSs during inspiral.
As we will show in \Sref{sec:inspiral}, the constraint violations diminish quickly 
during the evolution, the oscillations are small from the start, and the 
eccentricities are comparable.

In GR, it is difficult to quantify the NS spin, in particular for 
constraint-violating initial data. In a previous work \cite{Kastaun:2015:064027}, 
we used a volume integral formulation \cite{Gourgoulhon:2001:64029} of Arnowitt, Deser, Misner (ADM) angular 
momentum that is based on matter terms and has a similar form as the Newtonian formula. 
We then computed the difference between a spun-up model and the irrotational one.
For this work, we use the isolated horizon (IH) formalism 
\cite{Ashtekar:2003:104030, Ashtekar:2001:044016} instead.
We recall that this formalism is designed for measuring the spin of BHs using only 
spacetime quantities on the apparent horizon (see \cite{Ashtekar:2004:10} for a review).
A recent publication \cite{Tacik:2015:124012} has demonstrated that the IH
spin measure is also useful for NSs in binaries. 
Instead of apparent horizons, the formulas of the IH framework are applied to 
spherical surfaces around each of the stars.  
In the following, we will refer to the resulting measures as quasilocal (QL) 
spin.
Note that the method still does not involve 
the stress-energy tensor, but only the 4-metric.
Since our method of adding spin does not change the metric at all, the QL spin 
remains unchanged initially. The spins given in \Tref{tab:init_param} therefore 
refer to the time $0.5\msec$ after the start of the simulation. As will be shown in \Sref{sec:inspiral},
the spacetime has adapted to the rotating fluid at this time, i.e. the inconsistency 
described above (which is caused by constraint violations) has fallen to a tolerable level.
We report the QL angular momentum $J$ both as a rotation rate $F_R = J/(2\pi I)$ and as 
dimensionless value $\xi=J/M_g^2$, where $I$ and $M_g$ are moment 
of inertia and gravitational mass of the stars in isolation (for the SHT models, 
$I=48.45 \usk M_\odot^3$). For comparison, we also computed the difference of the 
ADM angular momentum for the SHT models, as in \cite{Kastaun:2015:064027}, and found 
values $\xi=\pm 0.119$, i.e. the two different measures agree within $5\%$.

\begin{table}[t]
	\caption{Initial data parameters. $M_{b}^\text{tot}$ is the total
          baryonic mass of the systems, $M_g$ is the gravitational
          mass of each star at infinite separation, and $d$ the initial proper
          separation.
          $F_R$ is the rotation rate of the stars with respect to the 
          irrotational case, and $\xi$ the corresponding dimensionless angular
          momentum. A positive sign means aligned with the orbital angular momentum,
          negative means antialigned.}
    \begin{ruledtabular}
	\begin{tabular}{@{}lllllll}
	Model             & EOS & $M_{b}^\text{tot}\,[M_\odot]$ & $M_g\,[M_\odot]$  &$d\,[\kilo\meter]$ 
	   & $F_R\,[\hertz]$ & $\xi$   \\\hline
	\texttt{SHT\_IRR} & SHT & $3.03$              & $1.40$           &$57.6$ & $0$             & $0$           \\
	\texttt{SHT\_UU}  & SHT & $3.03$              & $1.40$           &$57.6$ & ${+} 164$       & ${+} 0.125$   \\
	\texttt{SHT\_UD}  & SHT & $3.03$              & $1.40$           &$57.6$ & $\pm 164$       & $\pm 0.125$   \\
	\texttt{SHT\_DD}  & SHT & $3.03$              & $1.40$           &$57.6$ & ${-}164$        & ${-} 0.125$   \\			
	\texttt{APR4\_EM} & APR4 & $2.98$             & $1.35$           &$59.0$ & $0$             & $0$           \\
	\texttt{APR4\_UM} & APR4 & $3.01$             & $1.29,1.42$      &$59.0$ & $0,0$           & $0,0$           \\
	\end{tabular}
    \end{ruledtabular}
	\label{tab:init_param}
\end{table}

\subsection{Numerical evolution}
\label{sec:code}
For the numerical evolution of the spinning SHT models, we use the same numerical codes
as for the evolution of the irrotational SHT model, described in \cite{Kastaun:2016}.
The hydrodynamic evolution is performed by the \textsc{WhiskyThermal} code which 
employs finite-volume high resolution shock capturing methods in conjunction with
the HLLE approximate Riemann solver and the piecewise parabolic reconstruction method. 
We use a three-parameter tabulated EOS including thermal effects and 
composition in terms of the electron fraction (see \cite{Galeazzi:2013:64009} 
for technical details). 
We do not consider magnetic fields and neutrino radiation, and the electron fraction 
is passively advected along the fluid.
Vacuum regions are treated numerically by enforcing a minimum density (artificial atmosphere) 
of $6\times 10^7 \usk\gram\per\centi\meter\cubed$. Note this is a relatively high density
necessitated by the extent of the SHT EOS table in use. 
Regions covered by artificial 
atmosphere are not included in the stress energy tensor used as source term for the 
spacetime evolution.
During the inspiral, we also enforce adiabatic evolution, activating the full thermal
evolution shortly before the stars touch (see \cite{Kastaun:2016}).

The spacetime is evolved with the \texttt{McLachlan} code \cite{Brown:2009:44023},
which is part of the Einstein Toolkit \cite{Loeffler:2012:115001}. 
For all SHT models, spinning and irrotational, we chose the conformal and spatially 
covariant Z4 evolution scheme (CCZ4) described in \cite{Alic:2012:64040, Alic:2013:64049}, 
which has constraint damping capabilities that are advantageous when evolving 
constraint-violating initial data.
As gauge conditions, we use the $1+\log$-slicing condition \cite{Bona:1995:600} for 
the lapse function and the hyperbolic $\Gamma$-driver
condition \cite{Alcubierre:2003:84023} for the shift vector. At the outer boundary,
we use the Sommerfeld radiation boundary condition. To extract the multipole components
of the Weyl scalar $\Psi_4$, we use the modules \texttt{Multipole} and 
\texttt{WeylScal4} described in \cite{Loeffler:2012:115001}.
 
All codes are based on the Cactus Computational Toolkit infrastructure.
The time evolution of matter and spacetime is coupled using the Method of Lines (MoL) 
with fourth order Runge-Kutta time integration. We also use Berger-Oliger moving-box mesh 
refinement implemented by the \texttt{Carpet} code \cite{Schnetter:2004:1465}.
In detail, we use six refinement levels. During inspiral, the two finest ones consist 
of cubical boxes following the movement of the two stars (each completely contained 
in the corresponding smallest box). Shortly before merger, they are replaced by fixed 
levels centered around the origin, with an edge length of $60\usk\kilo\meter$ for 
the finest box. 
The finest grid spacing during the whole simulation is 
$295 \usk\meter$ and the outer boundary is located at $945\usk\kilo\meter$.
Finally, we use reflection symmetry across the orbital plane.
For tests of the code, we refer to \cite{Galeazzi:2013:64009,Alic:2013:64049}.

The two models with the APR4 EOS were evolved with a slightly different setup, 
as described in \cite{Ciolfi:2017:063016,Endrizzi:2016:164001}. The hydrodynamics was 
evolved using the \textsc{WhiskyMHD} code (with zero magnetic field) using a 
piecewise polytropic EOS, and without enforcing adiabatic evolution during inspiral.
The artificial atmosphere in this case had a lower density of 
$6\times 10^6\usk\gram\per\centi\meter\cubed$.
The spacetime was evolved with the McLachlan code, but using the Baumgarte, Shibata, Shapiro, Nakamura
(BSSN) formulation \cite{Nakamura:1987:1,Shibata:1995:5428,Baumgarte:1998:24007}. 
The outer boundary was located at $794$ and $1250\usk\kilo\meter$ for models 
\texttt{APR4\_UM} and \texttt{APR4\_EM}, respectively, and the finest resolution 
was $221\usk\meter$ for both.
In terms of stellar radius, this resolution is the same as for the less compact 
SHT models (the grid spacing is ${\approx}1/40$ of stellar equatorial coordinate 
radius or $1/50$ of circumferential radius). We will therefore use resolution 
studies carried out in \cite{Endrizzi:2016:164001,Ciolfi:2017:063016} for 
magnetized versions of the models above to estimate the numerical error for 
the SHT simulations. We caution that in \cite{Endrizzi:2016:164001,Ciolfi:2017:063016}
we could not demonstrate convergence for most quantities, likely because the lowest
resolution was too low, and estimated errors are based on the standard and high 
resolution runs only.

We also employ the diagnostic measure for the remnant properties introduced in 
\cite{Kastaun:2016}.
At regular time intervals during the evolution, we create 1D histograms of both
proper volume and baryonic mass contained in the grid cells, using bins of 
logarithmic rest-frame mass density. 
From the histograms, we then compute the total proper volume $V$ and the total baryonic 
mass $M_b$ contained inside isodensity surfaces.
This allows us to compute mass-volume relations that are still meaningful for the 
strongly deformed merger remnants and which are defined independent of the spatial 
gauge conditions. 
Note there is still a dependence on the time slicing, so when 
comparing different models the extrinsic curvature should be compared as well.
We expect the 1+log gauge used here to drive the system towards maximal 
slicing (the residual extrinsic curvature was not measured, however).
We also define a ``volumetric radius'' $R_V$ for the isodensity surfaces as 
the radius of a Euclidean sphere with equal volume, and from this a compactness 
measure $C=M_b/R_V$.
To define remnant properties without referring to some arbitrary density or radius cutoffs,
we use the notion of the bulk, defined as the interior of the isodensity surface
with maximum compactness $C$. Bulk mass and volume are then defined as baryonic mass
and proper volume of the bulk. For further details, see \cite{Kastaun:2016}.

For comparison, we computed the new bulk measures for sequences of stable 
TOV stars with the APR4 \cite{Akmal:1998:1804}, H4 \cite{Glendenning1991}, 
MS1 \cite{Mueller1996}, and SHT EOS \cite{Shen:2010:15806,Shen:2011:35802}, 
and $M_g$ between $1 \usk M_\odot$ and the maximum.
For those, the bulk (on $K=0$ time slices) contains 96.5\%--98.5\% of 
the total baryonic mass, thus motivating the name.
The new compactness is less than the standard compactness (gravitational 
mass divided by circumferential surface radius),
by a factor $0.85$--$0.93$. Bulk mass and compactness are
strictly monotonic functions of total baryonic mass and standard compactness, 
respectively.

\subsection{Improved GW extraction}
\label{sec:gw_extr}

To extract the GW signal, we use the standard approach of decomposing the Weyl scalar $\Psi_4$
into spin weighted spherical harmonics on a sphere with large radius close to the outer boundary
of the computational domain ($916$, $1181$, and $738 \usk\kilo\meter$ for the SHT, 
equal- and unequal-mass APR4 models, respectively). 
The tetrad choice for the computation of $\Psi_4$ is the one given in \cite{Loeffler:2012:115001}.
We do not extrapolate to 
infinity since we are only interested in a qualitative analysis.
To compute the strain, $\Psi_4$ has to be integrated in time twice, which will amplify any offset present
in the numerical waveform drastically. There are two standard methods in use to remove those 
offsets (see also the discussion in \cite{Maione:2016:175009}).
The most straightforward one is to fit a linear or quadratic function to the strain to determine 
the nonoscillatory part and then subtract it. However, we found this approach insufficient for our 
purposes. We are particularly interested in the behavior of the phase near regions where the strain 
amplitude temporarily becomes small. It is thus important that the offset is even smaller, which cannot 
be achieved by a simple fit since it will be dominated by the large amplitude parts.

The second common approach is the fixed frequency integration (FFI) method \cite{Reisswig:2011:195015}, 
where the integration is performed 
in frequency space by means of a Fourier transformation and the suppression of the low frequency part of the 
spectrum. This method
also turned out insufficient. The problem is that the Fourier spectrum of a linear drift has high frequency 
components because of the finite length of the waveform, overlapping the actual signal. In practice,
this causes strong drifts near the beginning and end of the resulting waveform.

To overcome these issues, we developed a new integration method applicable even to short waveforms.
The main idea is to first compute a local approximation to the integral based on the assumption of slowly
changing amplitude and frequency. The estimate is then subtracted from the numerically integrated function.
We then fit a slowly varying function to the difference, instead to the signal itself. Due to varying
frequency and amplitude, the oscillatory part does not cancel out completely when fitting  
to the latter. The key advantage of fitting the former is that
the residual oscillatory part is much smaller, thus reducing the impact on the drift correction.

In detail, the method works as follows. Given complex functions $z(t)=\dot{g}(t)$, we first compute the 
continuous phase $\phi_z$ such that $z(t) = z_a(t) e^{i\phi_z(t)}$, with $z_a \in \mathbb{R}$. 
Next, we compute a smoothed phase $\phi$ by convolution of $\phi_z$ with a Gaussian kernel of width $\sigma_s$,
where $\sigma_s \bar{\omega} = 4 \pi$, and $\bar{\omega}$ is the average phase velocity of $\phi_z$.
We require that $\phi$ is strictly increasing as a function of time (or strictly decreasing, 
we assume the former without loss of generality). Next, we express $z$ and $g$ as 
$z(\phi) = z_c(\phi) e^{i\phi}$ and $g(\phi) = g_c(\phi) e^{i\phi}$, and define $\omega\equiv\dot{\phi}$. 
The phases of the complex-valued 
amplitudes $z_c$ and $g_c$ change much more slowly than $\phi$ itself since we removed the main oscillation.  
By combining the equations $z(t)=\dot{g}(t)$ and $\dot{z}(t)=\ddot{g}(t)$, neglecting the second time 
derivative of $g_c$, and then rewriting time derivatives in terms of 
$\phi$-derivatives and $\omega$, we obtain an estimate $\hat{g}$ for $g$, given by
\begin{align}
\hat{g} &= \hat{g}_c e^{i\phi}, &
\hat{g}_c &= \frac{\frac{d z_c}{d\phi} - i z_c}{\omega + i \frac{d\omega}{d\phi}}.
\end{align}
We then compute the integral
\begin{align}
g(t_1) = \int_{t_0}^{t_1} z(t) \mathrm{d}t = \int_{\phi_0}^{\phi_1} \frac{z(\phi)}{\omega(\phi)} \mathrm{d}\phi
\end{align}
numerically (using trapezoidal integration).
Next, we define the residual $\delta g(\phi) = g(\phi) - \hat{g}(\phi)$.
In order to obtain the nonoscillatory part of $\delta g$, we fit a cubic spline $s$ with 
$N/m$ nodes regularly spaced in $\phi$, $N$ being the number of complete wave cycles in the signal.
Note that any oscillating contribution to $s$ will have frequencies lower by $1/m$ with respect to 
the actual signal. In this work, we chose $m=4$.
The final, drift-corrected result is simply $g-s$. 

We apply this integration method twice to obtain the GW strain $h$ from $\Psi_4$. 
Comparing the method to the aforementioned standard methods, we observed a strong reduction of 
artifacts near the boundaries, in particular for very short waveforms. We also found 
a much better alignment of low-amplitude parts of the signal. For our longest waveform, the 
phase velocity in the decaying tail remains usable longer than for the old methods,
although the phase velocity still becomes noisy when the amplitude becomes very small.
This seems to be partly because the signal is not dominated by a single mode anymore, 
rendering the phase velocity meaningless. In any case, the GW amplitude at this point is 
not relevant for GW astronomy anymore.

After a perfunctory comparison to strains obtained using an implementation 
\cite{Loeffler:2012:115001} of the Moncrief formalism 
\cite{Moncrief:1974:323} showed no obvious problems, we designed an analytic expression 
as a test signal for the new method in order to measure its accuracy. In detail,
\begin{align}
h(t) &= A(t) e^{i \phi(t)} + B t^2\\
A(t) &= \cos \left(\frac{3}{2} \pi \tanh \left(\frac{t}{\tau_1}\right)\right)  \\
\frac{1}{2\pi}\dot{\phi}(t) &= F(t) = F_1 + \frac{F_2}{1+e^{-t / \tau_2}} .
\end{align}
The amplitude goes to zero for $|t| \gg \tau_1$, and has three extrema, separated by 
two zero crossings. The instantaneous frequency increases from $F_1$ for $t \ll -\tau_2$ to 
$F_1 + F_2$ for $t \gg \tau_2$. Moreover, the second derivative of $h$ has an 
offset $2B$ in addition to the oscillatory part.
We picked parameters $B=300 \usk\second^{-2}$, $F_1=0.5 \usk\kilo\hertz$, 
$F_2 = 2 \usk\kilo\hertz$, $\tau_1 = 6 \usk\milli\second$, $\tau_2 = 4 \usk\milli\second$.
Next, we analytically computed the second derivative of $h$, sampled it on the 
interval $-15\ldots 15\usk\milli\second$, and applied the new integration 
method two times. Since the method is supposed to remove the nonoscillatory drift, 
we compare the result to the exact strain for $B=0$.
For this test, we find that the $L_1$-norm of the error of h is around $7\times 10^{-4}$ of 
the maximum amplitude, and $3\%$ of the $L_1$-norm of the offset term $B t^2$.
The $L_1$-norm of the phase error is around $0.002\usk\rad$.

For comparison, we also tested the FFI with a cutoff frequency of $F_1$.
We find that the error of $h$ is $6.5$ times larger than for the new method, 
and that the phase error is $18$ times larger, both mainly due to a large drift near the 
boundaries of the given time interval. Note however that the boundary effects affecting
the FFI method can be reduced by applying a window function before the Fourier 
transform, at the cost of significantly reduced usable signal length. 
We further note that also the polynomial fit method can be improved by applying a high-pass 
Butterworth filter, as shown in \cite{Maione:2016:175009, Feo:2017}. 
In general, it will depend on the use case which method is most suitable.

\subsection{Analyzing phase jumps in the GW signal}
\label{sec:jump_det}

While for BBH mergers the phase of the GW signal is relatively smooth, binary NS waveforms
often exhibit strong peaks in the phase velocity. Examples will be shown in \Sref{sec:gw}.
In some cases, they amount to a phase shift by $\pi$. 
Those phase jump events also seem to correlate with minima in the amplitude. This points
to a phenomenon called overmodulation in the context of signal analysis. Overmodulation
occurs for amplitude-modulated signals of the type
$z(t) = a(t) e^{i\omega t}$, when the modulation amplitude $a \in \mathbb{R}$ crosses zero. 
This corresponds to an instantaneous phase shift by $\pi$ and also leads to a nondifferentiable
signal amplitude $|a|$. 
An introduction to overmodulated signals and a method for their 
decomposition is given in \cite{Li:2004:5559172L}, which is however limited to real-valued signals.
We slightly generalize this idealized case by allowing $a$ to be 
complex valued, with a small imaginary part, and instead of the zero crossing we let $a$ pass 
close to the origin of the complex plane. The phase will then change rapidly and the amplitude 
$|a|$ will have a minimum 
where the second derivative is large. The larger the imaginary part, the larger the minimum 
amplitude and the slower the phase transition will occur.

Note that in the frequency domain, energy is removed from the carrier frequency and occurs as
sidebands for overmodulated signals. This is easy to understand, since parts of the signal
cancel each other out because their phases are inverted with respect to each other. It is therefore 
important to consider overmodulation in the context of GW astronomy, where a large part of the analysis is carried
out in frequency space. We will discuss the physical causes for overmodulation of GWs in 
\Sref{sec:gw}. In the following, we present an algorithm for the detection and removal 
of phase jumps. 

In order to detect possible phase jumps in a signal $z(t)$, we first compute a smoothed angular 
velocity $\omega$ from the phase velocity $\omega_z$ of $z$, using convolution with a Gaussian 
kernel of width $\sigma = 4\bar{P}$, where $\bar{P} = 2 \pi / \bar{\omega}$, and $\bar{\omega}$ is 
the time average of $\omega_z$.
We use the local maxima of $\eta = |\omega_z - \omega|$ as candidates for phase jumps. 
We ignore maxima below a cutoff $\eta < 0.1 \bar{\omega}$, 
and local maxima located closer than $2\bar{P}$ to a larger local maximum. 
We then try to model each remaining candidate as a phase jump event described by the 
generic function $\delta\phi_j(t) = \arctan\left(k\left(t-t_j\right)\right)$, which 
corresponds to an amplitude passing close to zero in the complex plane along a line 
with constant speed. The derivative is given by
\begin{align}
\omega_j(t) = \frac{d}{dt} \delta\phi_j(t) &= \frac{k_j}{1 + k_j^2 \left(t-t_j\right)^2}.
\end{align}
We then fit  $\omega_z(t)$ with the function 
\begin{align}
\omega_f(t) &= \omega_j(t) + k_{1j} + \left( t-t_j \right) k_{2j} ,
\end{align}
using fit parameters $k_j,k_{1j},k_{2j},t_j$.
Note $\omega_j$ has a very distinct shape near $t_j$, which sets it apart from other frequency 
fluctuations, e.g. caused by strong radial oscillations. It also decays rapidly away from 
$t_j$. We therefore limit 
the size of the fit interval $[t_j-\Delta t, t_j + \Delta t]$ to a certain fraction of the 
width of the peak given by $\omega_j$.
In terms of the phase shift $\alpha = \delta\phi_j(t_j+\Delta t) - \delta\phi_j(t_j - \Delta t)$
within the fit window, we chose $\Delta t$ such that $\alpha \approx \pi/4$.
Since $\delta\phi_j$ is unknown before the fit, 
we use an estimate computed from the area under the measured peak
\begin{align}
\alpha &\approx \int_{t_j-\Delta t}^{t_j+\Delta t} \left(\omega_z(t) - \omega_b(t)\right) dt ,
\end{align}
where $\omega_b$ models the background by linear interpolation of $\omega_z$ between the values 
at $t=t_j-2\bar{P}$ and $t_j+2\bar{P}$.
To avoid misinterpreting slow variations as jumps, we require that $\Delta t< 2\bar{P}$. 
As a last check, we also fit a simple quadratic function to the peak and ignore the candidate 
if the $L_2$ norm of the residual is less than the one for the fit with $\omega_f$.
The above heuristic method proved adequate to detect phase jumps for the GW strains
presented in this work robustly without manual intervention.

After fitting the individual phase jumps, we compute the combined phase correction as
$\delta\phi(t) = \sum_j \delta\phi_j(t)$.
From this, we define a complex amplitude 
$z_a(t) = |z(t)| e^{-i\delta\phi(t)}$. The real part will exhibit a zero crossing at the jumps,
while the imaginary part at the jump indicates how close to zero the signal gets.
We further obtain a corrected phase velocity 
$\omega = \frac{d}{dt} \left(\phi_z - \delta\phi \right)$. 
This phase velocity is a better indication for the frequency of the actual oscillations
of the source, since most of the contribution caused by the overmodulation
is removed. 
When applied to the test signal described in the previous section, we find that the 
corrected phase velocity agrees well with the exact value $\Omega(t)$. The $L_1$-norm
of the relative error is $0.2\%$. The largest deviations can be found near the jumps and 
near the boundaries, with a maximum error below 3\%.

\section{Results}
\label{sec:results}

In the following, we present the numerical results for the models shown in \Sref{sec:models}. 
The key quantities are summarized in \Tref{tab:outcome}.

\subsection{Inspiral}
\label{sec:inspiral}

\begin{table*}
\caption{Outcome of the mergers. 
  $M_\mathrm{blk}$, $R_\mathrm{blk}$, and $S_\mathrm{blk}$ are bulk mass,
  bulk radius, and bulk entropy,
  $\nu_\mathrm{max}$ denotes
  the remnants maximum rotation rate in the equatorial plane, 
  all computed $14\usk\milli\second$ after the merger.
  $f_\mathrm{pk}$ is the GW instantaneous frequency at merger time, 
  and $f_\mathrm{pm}$ is the frequency of the largest peak 
  in the postmerger part of the GW power spectrum.
  $M_\mathrm{disk}$ is the bound mass outside a coordinate sphere with
  proper volumetric radius $26$ km,
  $M_{60}$ is the bound mass outside $r>60$ km, both measured at 
  $t=14\, \mathrm{ms}$ after merger.
  Finally, $M_\mathrm{e}$ and $v_\infty$ are our best estimates for the total 
  mass of ejected matter and its average velocity at infinity.
  $M_\mathrm{e}^\mathrm{fix}$ is the ejected mass measured
  using a surface with fixed radius $222\usk\kilo\meter$.}
\begin{ruledtabular}
\begin{tabular}{lcccccc}
Model&
SHT Irr&
SHT UU&
SHT UD&
SHT DD&
APR4 UM&
APR4 EM
\\\hline 
$M_\mathrm{blk}\,[M_\odot]$&
$2.44$&
$2.46$&
$2.57$&
$2.44$&
---&
---
\\
$M_\mathrm{blk}/R_\mathrm{blk}$&
$0.220$&
$0.221$&
$0.223$&
$0.216$&
---&
---
\\
$S_\mathrm{blk}/M_\mathrm{blk} \,[k_\mathrm{B}/\mathrm{Baryon}]$&
$0.99$&
$0.84$&
$1.12$&
$1.29$&
---&
---
\\
$\nu_\mathrm{max}\, [\mathrm{kHz}]$&
$0.95$&
$0.94$&
$0.96$&
$0.95$&
$1.64$&
$1.66$
\\
$f_\mathrm{pk}\, [\mathrm{kHz}]$&
$1.42$&
$1.45$&
$1.33$&
$1.41$&
$2.09$&
$2.12$
\\
$f_\mathrm{pm}\, [\mathrm{kHz}]$&
$2.02$&
$2.09$&
$2.18$&
$1.96$&
$3.30$&
$3.47$
\\
$M_\mathrm{disk}\,[M_\odot]$&
$0.291$&
$0.296$&
$0.163$&
$0.236$&
---&
---
\\
$M_{60}\,[M_\odot]$&
$0.158$&
$0.159$&
$0.053$&
$0.106$&
---&
---
\\
$M_\mathrm{e}\, [M_\odot]$&
$0.0003$&
$0.0011$&
$<10^{-4}$&
$0.0007$&
$0.0100$&
$0.0126$
\\
$M_\mathrm{e}^\mathrm{fix}\, [M_\odot]$&
$<10^{-4}$&
$0.0009$&
$<10^{-4}$&
$0.0003$&
---&
---
\\
$v_\infty\, [c]$&
$0.10$&
$0.12$&
$0.06$&
$0.10$&
---&
---
\end{tabular}
\end{ruledtabular}
\label{tab:outcome}
\end{table*}

\begin{figure}
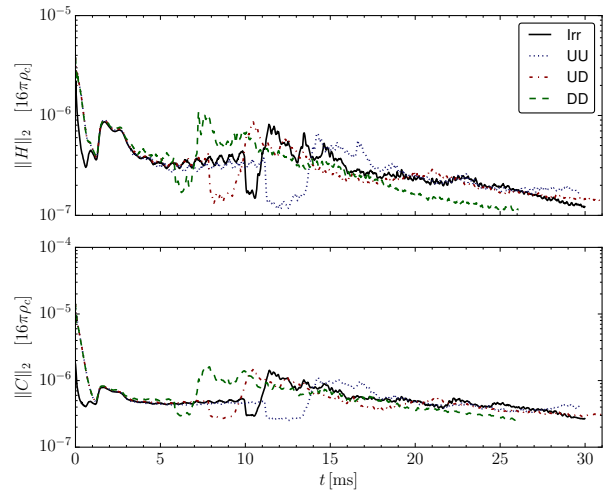

  \begin{center}
    \includegraphics[width=0.95\columnwidth]{{{constraints}}}  
    \caption{Evolution of constraint violations for the SHT EOS models. 
    Top panel: $L_2$-norm $\Vert H \Vert_2$ of the 
    Hamiltonian constraint $H$. Bottom panel: combined $L_2$ norm 
    $\Vert C \Vert_2 \equiv (\sum_{i=1}^3 (\Vert C_i\Vert_2)^2)^{1/2}$ of momentum constraints $C_i$. 
    All constraints are normalized using the central density $\rho_c$ of the initial stars.}
    \label{fig:constraints}
  \end{center}
\end{figure}

We start by assessing the potential problems with our spinning initial data described
in \Sref{sec:models}.
The constraint violations for the different spin configurations are shown in 
\Fref{fig:constraints}. 
For the spin values considered here, the initial Hamiltonian
constraint is only larger by a factor around $2.5$ compared to the irrotational model,
and the combined momentum constraint by a factor $8$.
Thanks to the CCZ4 evolution formalism, Hamiltonian and momentum constraint violations
for the spinning models further decrease to the level of the irrotational model within 
$1\usk\milli\second$. In the later inspiral, the constraint violations as a function of time 
differ again, but only because there is a physical impact on the duration of the inspiral,
as will be discussed later. The amount of constraint violation at merger time is very 
similar for the different spins.

\begin{figure}
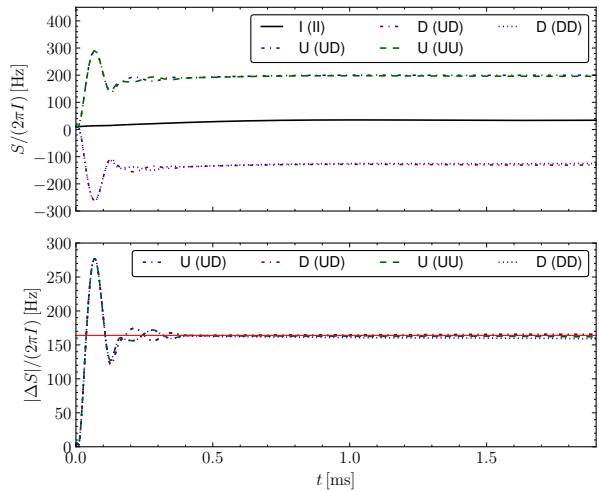

  \begin{center}
    \includegraphics[width=0.95\columnwidth]{{{ih_spins}}}  
    \caption{Top panel: 
    Quasilocal angular momentum $S$ of the individual stars for the SHT EOS models.
    $I$ is 
    the moment of inertia of the TOV solution with the same baryonic mass.
    Stars with aligned, antialigned, and irrotational spin configurations are denoted    
    U, D, and I, respectively, with the binary model they are part of given in brackets. 
    Bottom panel: The difference of the spin to the one for the irrotational 
    model. The horizontal line marks the nominal spin of $164\usk\hertz$. }
    \label{fig:ih_spins}
  \end{center}
\end{figure}

Satisfying the constraints only means that the spacetime is consistent on the
corresponding time slice; it does not imply that it represents the initial data we want, 
i.e. spinning NSs in quasicircular orbit. 
In order to measure how the metric near the stars adapted to the manually added NS 
rotation, we use the quasilocal spin described in \Sref{sec:models}.
Figure~\ref{fig:ih_spins} shows the QL angular momentum for the first 
$2\usk\milli\second$ of the simulations, extracted on spherical 
surfaces with radius $13.3\usk\kilo\meter$ around the 
NS barycenters. 
Since we do not change the metric at all, the curves for the spinning models start at 
the same value as the irrotational model. Within less than $0.5\usk\milli\second$,
however, the spacetime near the stars apparently adjusts to the rotation of the fluid.
We thus regard the system at this time as our true initial data, and the spin 
values reported in \Tref{tab:init_param} are the differences of the QL measure to 
the irrotational model at $0.5\usk\milli\second$.

Note that the QL spin for the irrotational model is not exactly zero initially, we find 
$\xi\approx 0.008$. Although there is no reason to expect it to be zero for actual 
irrotational binaries at finite distance, it is unknown how much of the residual spin 
is generated by the imperfections of the irrotational initial data. Using a different 
initial data code for a different model, but with comparable initial separation, 
\cite{Tacik:2015:124012} reported a much smaller value of $\xi = 2 \times 10^{-4}$.

Also note that the QL spin measure exhibits a slow drift during the whole inspiral. 
This drift seems to be the same regardless of the NS spin, even for the irrotational model.
The reason might be that it becomes more difficult to separate orbital and individual 
spin contributions to the metric when the stars come closer. This is however beyond the 
scope of this work; the issue is investigated in more detail in \cite{Dietrich:2016:044045}.
Here we only used the QL spin to assess the quality of the initial data.
In this regard, we also note that the NS oscillations induced by adding the spin are rather 
small for our cases. The central density varies by less than $0.6\%$ for all spin 
configurations. For comparison, the central density oscillation amplitude for the 
irrotational model is around $0.2\%$.

\begin{figure}
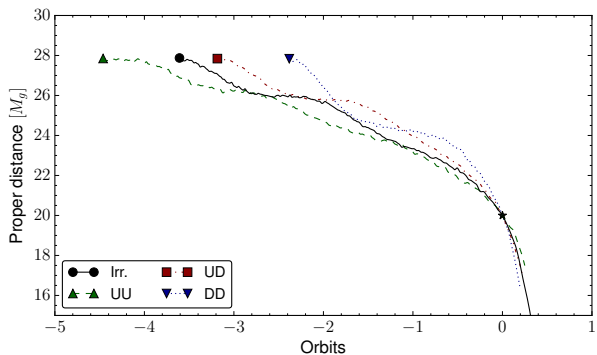

  \begin{center}
    \includegraphics[width=0.95\columnwidth]{{{separation}}}  
    \caption{Evolution of proper separation between the NSs density maxima versus
      orbital phase, for the SHT EOS models. The separation is given in units of the single star gravitational 
      mass in isolation, and the curves have been aligned at a separation of $20 M_g$}
    \label{fig:separation}
  \end{center}
\end{figure}

Figure~\ref{fig:separation} shows the proper separation between
the density maxima of the NSs versus their orbital phase (measured with respect to 
simulation coordinates). Judging by eye, both the irrotational and the spinning models show 
a similar degree of eccentricity. We therefore assume the additional error in the linear 
momentum due to spin is unimportant, but caution that a solid error estimate would require
further simulations and that, strictly speaking, all our results are valid for a slightly 
eccentric binary.

The number of orbits until merger however is clearly affected by the initial NS spin:
increasing the spin of the NSs aligned with the orbital angular momentum prolongs 
the inspiral, while in the antialigned case it is accelerated.
The inspiral time (time between the start of simulation and merger) follows the same trend 
as the number of orbits, with $7.4$, $10.0$, $11.3$,  and $14.0 \usk\milli\second$ for 
the down-down, up-down, irrotational, and up-up configurations, respectively.
To estimate the error due to finite resolution, we use the resolution study in 
\cite{Endrizzi:2016:164001}.
For this case, second order convergence was demonstrated for the GW signal during inspiral,
and we estimate a total phase error during six orbits of inspiral of around $0.6\usk\rad$. 
However, this just constitutes a best guess, since the resolution study was done for an 
irrotational unequal-mass model with a different EOS (APR4).
Other differences are likely less important:
the use of the BSSN formalism instead of CCZ4 should result in comparable errors, as 
demonstrated in \cite{Alic:2013:64049};
although \cite{Endrizzi:2016:164001} evolved the magnetohydrodynamic evolution equations 
as well, this will, if anything, increase the error;
the lower density of the artificial atmosphere is not relevant for the inspiral since 
the (Newtonian) dynamic pressure obtained for the SHT simulations in this work corresponds
to deceleration time scales $v/\dot{v} \gtrsim 10^3 \usk\second$.

Our findings further corroborate this orbital hangup effect, which was already described
in previous studies \cite{Tsatsin:2013:64060, Bernuzzi:2014:104021, Kastaun:2015:064027} 
for different models and, more importantly, three different methods of adding the initial 
NS spin. In order to quantify the effect precisely, it would be necessary to further
reduce the eccentricity, using e.g. the methods described in 
\cite{Kyutoku:2014:064006,Dietrich:2015:124007,Tacik:2015:124012}.
As a ballpark figure for the cases at hand, varying the initial spin between $\pm 164\usk\hertz$ 
changes the length of the inspiral by two full orbits.

\subsection{Postmerger dynamics}  
\label{sec:vortexevol}
To get an overview of the fluid flow during and after merger, we computed
the fluid trajectories in the equatorial plane in a coordinate system corotating 
with the $m=2$ component of the density perturbation in the equatorial plane. The 
latter is computed from the phase of the complex-valued $m=2$ moment of the density in 
the equatorial plane. Since we only want to remove the average rotation for visualization
purposes, but not introduce noise in the trajectories, we smoothed the phase in time using 
convolution with a $2\usk\milli\second$ Hanning window. The trajectories are computed
by integrating backwards in time in a postprocessing step, as described in 
\cite{Kastaun:2016}. In particular, we use the coordinate system defined there to
suppress deformations due to gauge effects.

\begin{figure*}
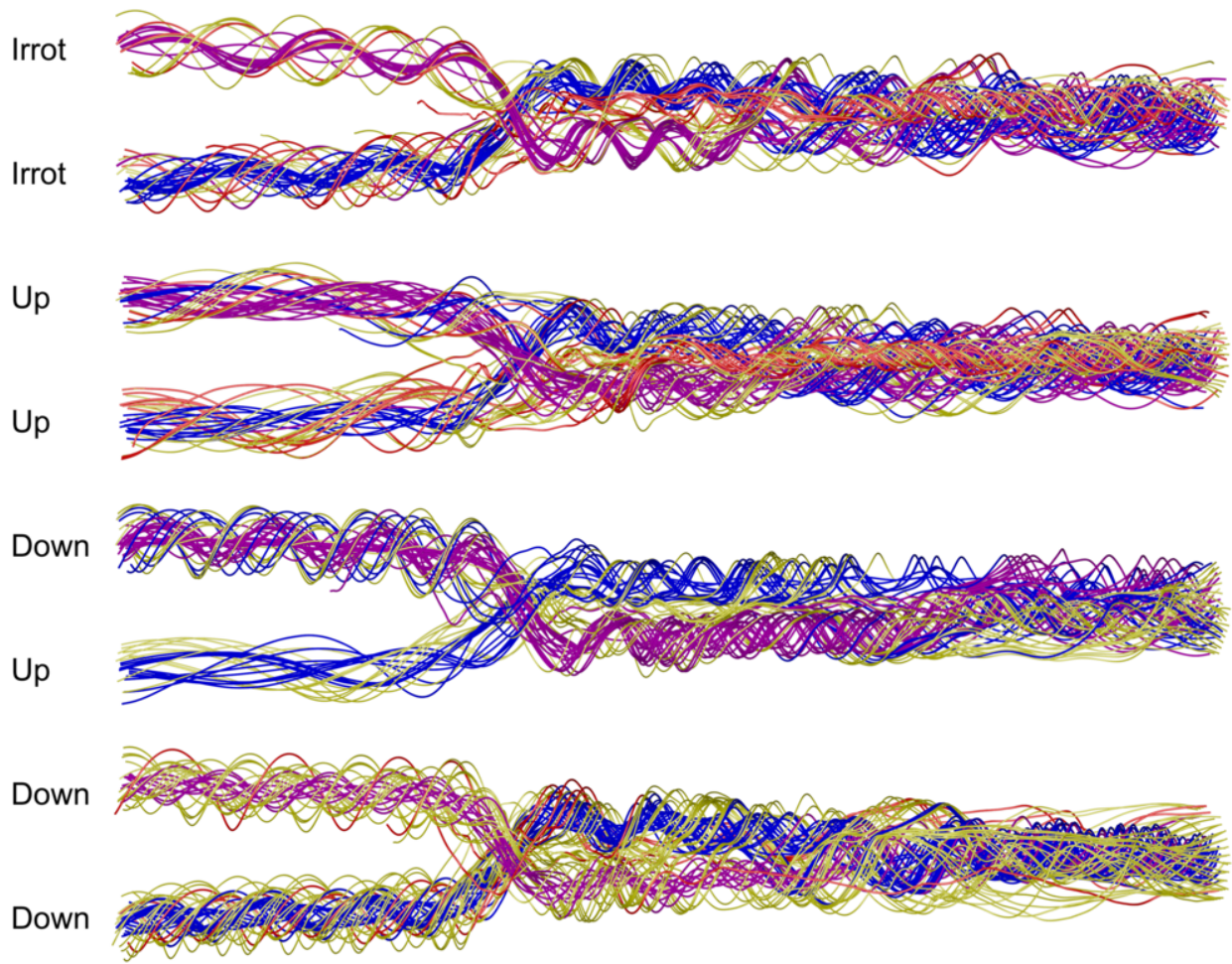

  \begin{center}
    \includegraphics[width=0.95\textwidth]{{{sht_traj_xyt_unscrewed}}}  
    \caption{Comparison of fluid trajectories in the equatorial plane for different initial spins, for the SHT EOS.
    The time runs from left to right
    and spans the interval $7\usk\milli\second$ before merger to $14\usk\milli\second$ after merger.
    The coordinates in the equatorial plane are corotating with the density deformation (see text);
    the larger principal axis ($x$ axis) is pointing up. 
    Only trajectories remaining in the remnant are shown, and a density cutoff 
    at $10^9 \usk\gram\per\centi\meter\cubed$ is applied 
    (this is the reason why some trajectories start later). 
    Trajectories belonging to the 
    two main vortices after merger are colored blue and violet, trajectories passing a secondary vortex in the 
    early and late postmerger phases are colored red and yellow, respectively. }
    \label{fig:traj_xyt}
  \end{center}
\end{figure*}

In \Fref{fig:traj_xyt}, we show the trajectories as a 2+1D spacetime diagram where 
two dimensions correspond to the equatorial plane.  Here we compare the results for all
spin configurations to the irrotational case (which was shown in a similar plot
already in~\cite{Kastaun:2016}). 
Removing the overall rotation is necessary in order to visualize the vortex structure
of the fluid flow. We note however that there is a certain ambiguity in the choice 
of the rotation since during merger, the principal axes of the $m=2$ deformation
are exchanged. This leads to a rotation of the plot coordinates by $90\degree$ 
within the smoothing length, which is advantageous for the purpose of the 
visualization. In the following, we refer to the larger and smaller principal axes 
of the remnant $m=2$ deformation in the equatorial plane as $x$ and $y$ axes.

Before we discuss the differences, we review the features common to all models.
Directly after merger, the cores of the NSs have formed two large vortices which are 
rotating against each other. Their separation vector is oriented along the $x$ axis.  
This continues for several milliseconds until the two vortices
slowly merge into a single larger vortex that is better described as differential rotation,
but with a strong nonaxisymmetric deformation.
In addition to the main vortices, we also observe smaller secondary vortices, which 
become more elongated over time, but are still present at the end of the simulation.
Those secondary vortices are located in the half-planes $y>0$ and $y<0$, i.e. the centers 
are rotated around $90\degree$ with respect to the density distribution. 
To highlight the primary vortices in the figure, we colored trajectories blue or purple
if they stay in the $x<0$ or $x>0$ half-planes, respectively, during the early post 
merger phase ($1$--$5\usk\milli\second$ after merger). Trajectories which stay in one 
of the half-planes $y>0$ or $y<0$ during the early (late) postmerger phase are 
colored red (yellow), to highlight the secondary vortices. We note that some trajectories 
also escape or join the vortices.
All vortices are phase locked with the main density perturbation.

Note that directly after merger, we also expect a Kelvin-Helmholtz (KH) instability, 
which is however not well resolved at the resolution used here. A high-resolution 
numerical study for the case of magnetized systems can be found in \cite{Kiuchi:2015}.
For our simulations, we expect that the time scale on which
the  two vortices merge is most likely affected, but it is difficult to estimate 
the numerical error.

As can be seen in \Fref{fig:traj_xyt}, the general evolution outlined above is strongly 
influenced by the initial spin. The first difference is that the 
large vortices rotate with different speeds against each other, which is to be expected.
Figure~\ref{fig:traj_xyt} also provides a first indication that the size and shapes of the 
secondary vortices are strongly affected. We will come back to this later on.
It is also worth noting that the mixed spin model \texttt{SHT\_UD} shows a strong asymmetry,
apparently related to the fact that the two main vortices rotate with different speed.

Close inspection of \Fref{fig:traj_xyt} reveals an unequal number of  trajectories in 
the two inspiraling NSs even for the $\pi$-symmetric models. This is unexpected since 
the seed positions at the end of the postmerger phase are also $\pi$-symmetric. 
Note however that the mapping from the trajectory positions at the end of the 
simulation to the positions at the beginning is not smooth at all. The constant
churning motion and the crossing of trajectories from one vortex to another is 
likely to make this map almost chaotic in the sense that small changes in the initial
conditions lead to large changes in the outcome. Under perfect conditions, this 
still cannot break the symmetry. However, numerical errors and possibly physical
instabilities will introduce small asymmetries. Considering also the aforementioned 
selection rules for the plotted trajectories, it is hard to predict what distribution
of trajectories in the inspiral phase will result even from small asymmetries. 
We therefore caution that \Fref{fig:traj_xyt}  
should be used only to visualize the overall fluid flow.

The deformation of the remnant, its thermal structure and flow patterns $8\usk\milli\second$ after merger 
are depicted in \Fref{fig:vortex_struct_t8} (similar movies are available in the supplemental material 
\cite{Kastaun:2017:supplemental}).
At the time shown in the figure, the main vortices have already merged for all models except \texttt{SHT\_UD}. 
Using movies corresponding to \Fref{fig:vortex_struct_t8}, we found that the pattern shown in 
\Fref{fig:vortex_struct_t8} is quasistationary on time scales of several milliseconds. Since we 
use a frame corotating with the pattern, this corresponds to the existence of an approximate helical 
Killing vector. In this sense, the remnant is not highly dynamic despite a strong nonlinear deformation.

Clearly, there is a relation between secondary vortices, hot spots, and density perturbation, although 
cause and effect are uncertain. As discussed for the irrotational model in \cite{Kastaun:2016}, 
the hot spots apparently consist partially of hot matter trapped inside the secondary vortices, but also
of regions where the temperature is raised by adiabatic heating at a local compression of the fluid flow. 
To substantiate that statement, we sample temperature and specific entropy along fluid trajectories 
and plot them as a function of the $\phi$ coordinate in the frame corotating with the density pattern, as
shown in \Fref{fig:traj_thermal_irr} for the irrotational model. One can see that compared to the 
temperature, the specific entropy shows only small variations along the trajectories (note that any 
decrease has to be caused by numeric errors such as heat conduction by numeric dissipation or the errors in the 
trajectory tracing).
It is also apparent that some trajectories stay in the vortices, others pass through, and some stay a few
cycles before leaving. The trajectories passing through show a strong adiabatic temperature increase
in the regions rotated $\pm 90\degree$ to  the density pattern, and the material trapped in the hot spots
has on average a somewhat larger specific entropy.
Likely, both the thermal pressure and the varying centrifugal forces in the secondary vortices  
influence the density perturbation. The density perturbation, in turn, limits the possible fluid flows for 
a quasistationary pattern.

Our findings indicate that the early postmerger phase is not highly dynamic, but nevertheless very complex. 
It is doubtful whether it can be described correctly by linear mode analysis of
axisymmetric stationary background models. 
This should be kept in mind when interpreting studies that use normal mode terminology
to describe merger remnant oscillations, 
e.g. \cite{Stergioulas:2011:427, Bernuzzi:2014:104021,Foucart:2016:044019} to 
name just a few (in fact \cite{Stergioulas:2011:427} already cautions about 
the lack of a complete 
linear normal-mode analysis of merger remnants).
In particular, interpreting the main frequency peak 
from the postmerger signal as the frequency of a linear $l=m=2$ $f$ mode seems questionable.
A possible alternative would be to use the assumption of a helical Killing vector to compute strongly 
nonaxisymmetric systems. Modeling this phase correctly is important for the interpretation of 
future GW detections since only the early postmerger phase might be detectable, given the decaying 
amplitude. 

Figure~\ref{fig:vortex_struct_t8} shows a strong influence of the spin on the early remnant. Both the size and 
the shape of the secondary vortices differ, and so does the density perturbation. For the mixed spin case
\texttt{SHT\_UD}, the two main vortices merge later than for the 
other cases, and are also asymmetric. Moreover, there is only one secondary vortex.
Finally, we notice that the secondary vortices seem to be involved in transporting 
matter out of the remnant into the disk (compare Fig.~17 in \cite{Kastaun:2016} and the 
related discussion). Also this aspect is influenced by the spin.
Assuming that our findings also apply to hypermassive stars, this might be relevant in the context 
of short gamma ray bursts, where the disk mass after collapse to a BH is a key parameter.

The differences in the structure are also reflected in the compactness of the remnant.
Figure~\ref{fig:bulk_comp} shows the evolution of the bulk compactness introduced in \Sref{sec:code}. 
There are significant differences in the postmerger phase, up to $5\%$. Those persist on the 
time scale of our simulations; the value at time $14\usk\milli\second$ after merger is given in 
\Tref{tab:outcome} (throughout this work, time of merger refers to the retarded time of the 
first maximum of the GW strain amplitude). 

\begin{figure*}
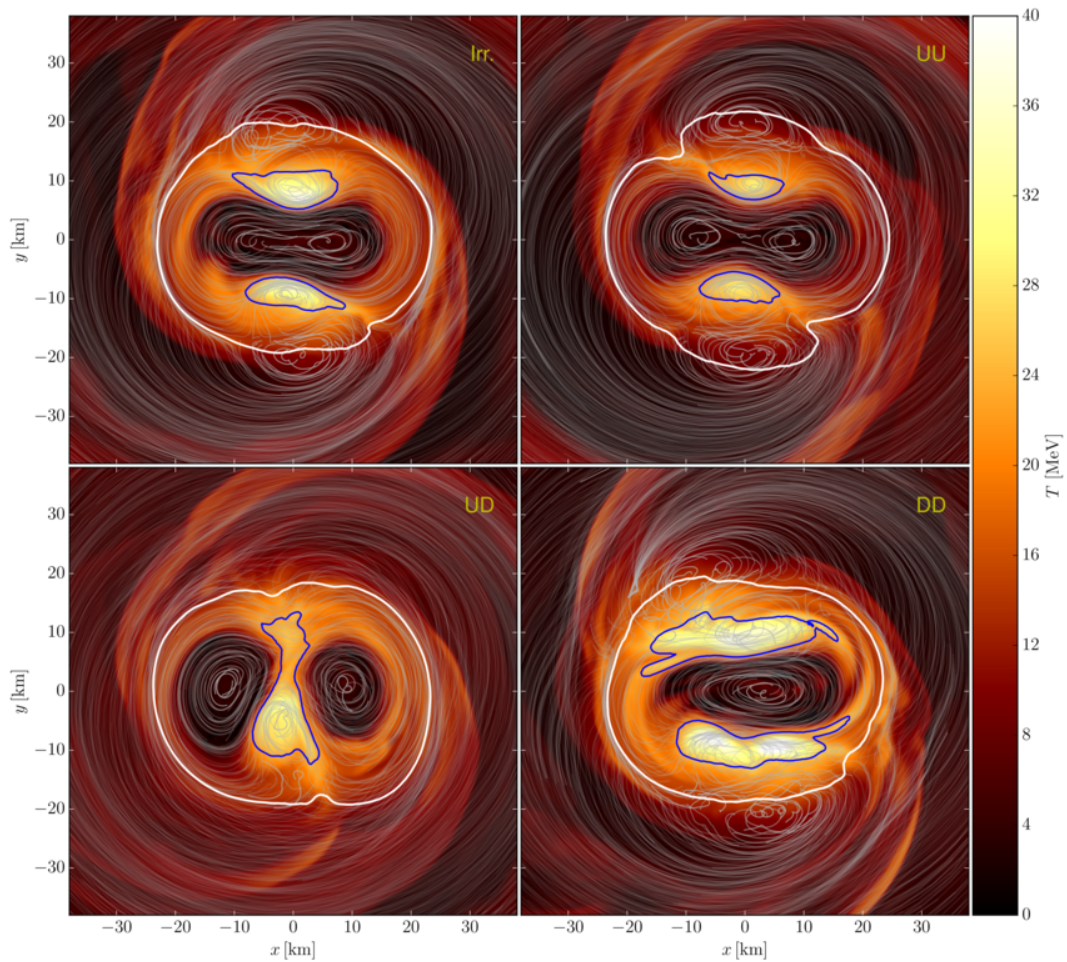

  \begin{center}
    \includegraphics[width=0.95\textwidth]{{{temp_traj_xy_t8}}}  
    \caption{Comparison of vortex structure $8\usk\milli\second$ after merger for different initial spin.
    The temperature is shown as color plot. The thick white line is the isodensity contour corresponding 
    to the bulk. The blue lines are isocontours of entropy density, marking the hot spots. The thin white lines
    are fluid trajectories in the frame corotating with the $m=2$ density perturbation, in a time window 
    from 7 to 9$\usk\milli\second$ after merger.}
    \label{fig:vortex_struct_t8}
  \end{center}
\end{figure*}

\begin{figure}
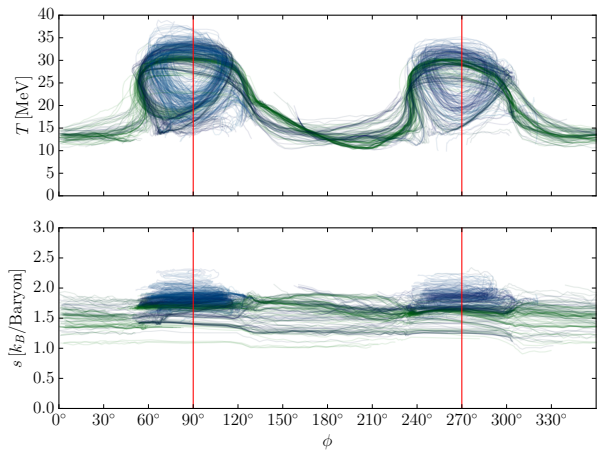

  \begin{center}
    \includegraphics[width=0.95\columnwidth]{{{traj_thermal_Irr}}}  
    \caption{Temperature (top panel) and specific entropy (bottom panel) along fluid trajectories as 
    a function of the $\phi$ coordinate in the frame corotating with the density perturbation, for model 
    \texttt{SHT\_IRR}, in the time interval $5$--$11 \usk\milli\second$ after merger.
    Shown are only trajectories related to the hot spots,
    selected by the requirement that the maximum temperature is at least $30\usk\mega\electronvolt$ and that 
    the minimum density is above the bulk density. The monotonicity of $\phi$, defined as
    $\mu = (\int \dot{\phi})/(\int|\dot{\phi}|)$, is denoted by colors ranging from green for $\mu=-1$ to blue
    for $\mu=0$. }
    \label{fig:traj_thermal_irr}
  \end{center}
\end{figure}

\begin{figure}
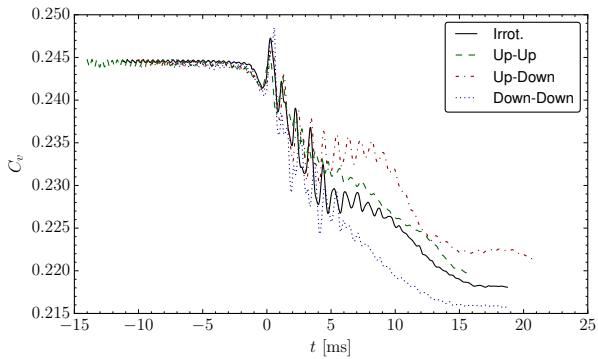

  \begin{center}
    \includegraphics[width=0.95\columnwidth]{{{bulk_comp}}}  
    \caption{Evolution of bulk compactness for different initial spins. The time $t$ is relative 
    to the time of merger.}
    \label{fig:bulk_comp}
  \end{center}
\end{figure}

\subsection{Late remnant structure}
\label{sec:rotprof}
We now discuss the structure of the remnant in a later evolution stage, 
around $14\usk\milli\second$ after merger.
As shown in \Fref{fig:vortex_struct_t14}, the system has become more axisymmetric,
in particular the hot spots have become ring shaped. The secondary vortices are 
still present, but have become less localized.
This might be a consequence of the reduced density perturbation, since the early vortices
are situated in the wakes of the most radially extended parts.
Note that the mixed spin model \texttt{SHT\_UD} is still very asymmetric, but the pattern 
changed drastically with respect to the earlier state shown in 
\Fref{fig:vortex_struct_t8}. The system underwent a rearrangement of vortices within
$2$--$3\usk\milli\second$. In contrast, the changes of the other models took place 
more gradually.

\begin{figure*}
  \begin{center}
    \includegraphics[width=0.95\textwidth]{{{temp_traj_xy_t14}}}  
    \caption{Like \Fref{fig:vortex_struct_t8}, but showing the system 
    $14\usk\milli\second$ after merger.}
    \label{fig:vortex_struct_t14}
  \end{center}
\end{figure*}

\begin{figure*}
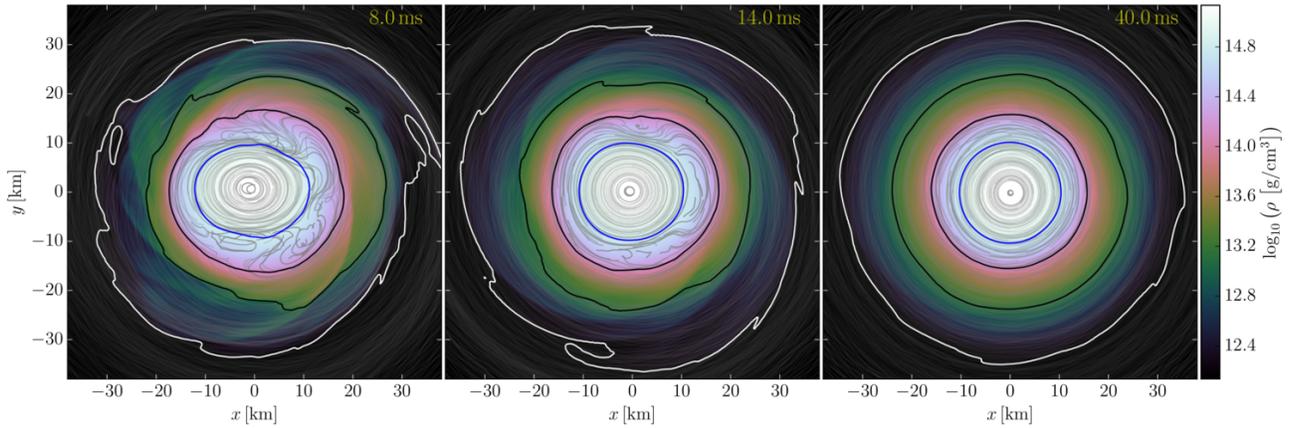

  \begin{center}
    \includegraphics[width=0.95\textwidth]{{{apr4_em_traj_xy}}}  
    \caption{Comparison of vortex structure in the equatorial plane for model \texttt{APR4\_EM} at 
    times $8,14$ and $40\usk\milli\second$ after merger. The density is shown as a color scale.
    The thick lines are isodensity contours $0.5, 0.1, 0.01, 0.001$ of the maximum density. 
    The thin grey lines
    are fluid trajectories in the frame corotating with the $m=2$ density perturbation, in a time window 
    $\pm 1\usk\milli\second$ around the snapshot.}
    \label{fig:vortex_evol_apr4_em}
  \end{center}
\end{figure*}

To rule out that the remnant structure described so far is specific to the corner case of
stable remnants of equal-mass binaries, we also investigate the vortex structure for the two supramassive models,
\texttt{APR4\_EM} and \texttt{APR4\_UM}, one of which is an unequal mass model. Figures~\ref{fig:vortex_evol_apr4_em}
and \ref{fig:vortex_evol_apr4_um} show the trajectories on top of the mass density (the temperature was not available 
for those runs). Again, we find secondary vortices phase locked with the main density perturbation, although they
are less pronounced for the equal mass model. Not surprisingly, the unequal mass model has a strongly asymmetric 
vortex structure. To a lesser degree, also the equal mass system deviates from $\pi$-symmetry. The asymmetry 
correlates with the vortices and might be caused by their dynamics in the early postmerger phase. In any case, 
the system becomes more axisymmetric with time, which rules out an unstable $m=1$ mode as the cause.
 
\begin{figure*}
  \begin{center}
    \includegraphics[width=0.95\textwidth]{{{apr4_um_traj_xy}}}  
    \caption{Like \Fref{fig:vortex_evol_apr4_em}, but showing model \texttt{APR4\_UM} at 
    times $8$ and $14\usk\milli\second$ after merger.}
    \label{fig:vortex_evol_apr4_um}
  \end{center}
\end{figure*}

In contrast to earlier stages, the remnants at the time shown in \Fref{fig:vortex_struct_t14}
can be described roughly as differentially rotating objects, albeit still strongly 
nonaxisymmetric and with superposed vortices. It therefore becomes meaningful to measure
the rotation profile. To this end, we use two methods. First, we compute the angular velocity 
averaged in the $\phi$ direction at fixed radii. Second, we average the angular velocity and the 
radial coordinate of fluid elements along their trajectories over a fixed time interval. 

\begin{figure*}
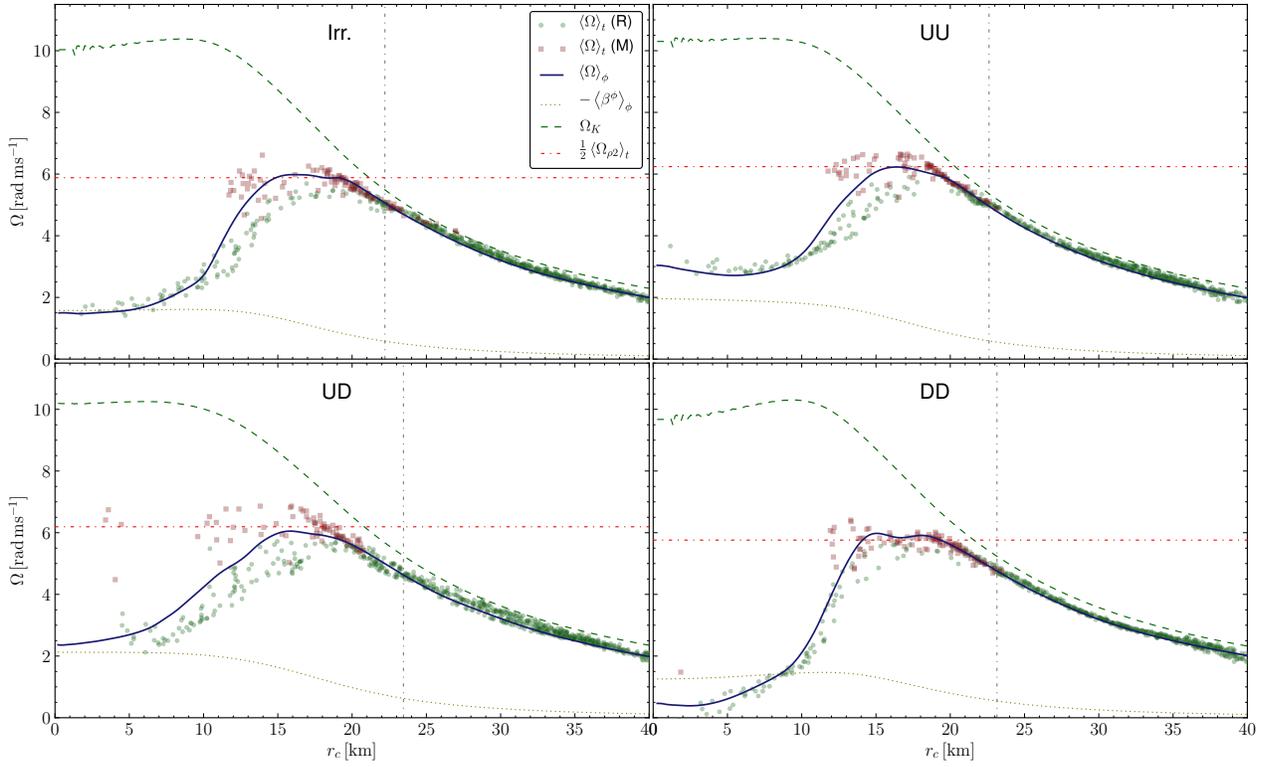

  \begin{center}
    \includegraphics[width=0.95\textwidth]{{{rot_prof_t14}}}  
    \caption{Angular velocity in the equatorial plane as observed from infinity 
    at $14\usk\milli\second$ after merger, averaged over a time interval 
    $\pm 1\usk\milli\second$. 
    The solid line shows the $\phi$ average of angular velocity.
    The markers are the time average of angular velocity and circumferential
    radius along traced fluid trajectories.
    Trajectories of mixed prograde and retrograde nature (with respect to the corotating 
    frame) are marked by red squares, the others by green circles.
    The vertical lines mark the radius at which the average density is 5\% of the central 
    one, and the horizontal lines the angular velocity of the $m=2$ component
    of the density perturbation in the equatorial plane. 
    The dashed curve is the estimated orbital angular velocity of a test mass in corotating 
    orbit. The dotted curve is the $\phi$-averaged shift vector component $\beta^\phi$
    as an estimate for the frame dragging coefficient.
    }
    \label{fig:rot_prof_t14}
  \end{center}
\end{figure*}

The results are shown in \Fref{fig:rot_prof_t14}. 
Not surprisingly, the trajectories making up the secondary vortices show a large spread.
On average, however, the angular velocity of those trajectories coincides both with the 
maximum average in the $\phi$ direction and with the angular velocity of the density 
perturbation (shown as horizontal lines in \Fref{fig:rot_prof_t14}). Apparently, fluid 
elements cannot overtake the elevated parts of the deformation pattern.
Since the density deformation is responsible for the GW signal, this explains our 
observation that the maximum rotation rate is 
approximately given by 
half the GW frequency.
As shown in \Tref{tab:outcome}, the difference between maximum rotation at 
$14\usk\milli\second$ after merger and the main postmerger peak in the GW spectrum is 
between $0.6\%$ and $14\%$ for our models. 
Since the frequencies are evolving, it is more meaningful to compare the 
time-dependent maximum rotation rate to half the instantaneous GW phase 
velocity (cf. Fig.~15 in \cite{Ciolfi:2017:063016}). For this, we considered 
times $>5 \usk\milli\second$ after merger, except for the up-down model, 
where the maximum rotation rate around the origin is still the central one 
until ${\approx}10 \usk\milli\second$ after merger (due to the off-center 
vortex, see \Fref{fig:vortex_struct_t14}). We also ignored times after 
$25 \usk\milli\second$ for the \texttt{APR4\_EM} model because of the low 
GW amplitude. On those intervals, the $L_1$-norm of the difference is 
$\lesssim 6\%$ for all models. The relation between maximum rotation rate 
and GW frequency was also observed in
\cite{Kastaun:2015:064027,Endrizzi:2016:164001, Hanauske:2016}.
We also find that the remnant cores rotate slowly. The same was also found 
in \cite{Kastaun:2015:064027,Endrizzi:2016:164001, Hanauske:2016} for various hypermassive and 
supermassive remnants with different EOSs, and in \cite{Kastaun:2016} for our irrotational 
model. 
Note that the main vortex for the mixed-spin model is displaced by a secondary vortex (see \Fref{fig:vortex_struct_t14}), 
which explains the trajectories with high angular velocity at $r<5\usk\kilo\meter$ visible in 
\Fref{fig:rot_prof_t14}. As for unequal mass models, the central rotation rate is not 
meaningful in this case since the center is not the center of rotation.

To estimate the error of the rotation rates, we use the maximum rotation 
rates reported in \cite{Ciolfi:2017:063016}. 
Although the resolution test in there was done for a magnetized version, the influence of
the magnetic field on the rotation profile was shown to be small.
Under the optimistic assumption 
of first order convergence starting at the standard resolution, we estimate 
the error of the rotation rates around $3\%$.
The impact of the initial spin on the maximum rotation rate reported in \Tref{tab:outcome}
is only around $2\%$. The amount of differential rotation shown in \Fref{fig:rot_prof_t14}
on the other hand exhibits larger differences, exceeding the numerical error.
We note there is a possibility that the differential rotation might in 
reality be reduced quickly by viscous effects induced by strong 
magnetic fields on small length scales. Those can be generated in principle 
by KH and magnetorotational instabilities that are not resolved in numerical 
simulations with standard resolutions (see \cite{Kiuchi:2015}). The possible 
effect of such viscous damping was recently pointed out by 
\cite{Shibata:2017:123003} using a simple alpha-viscosity model.

\begin{figure}
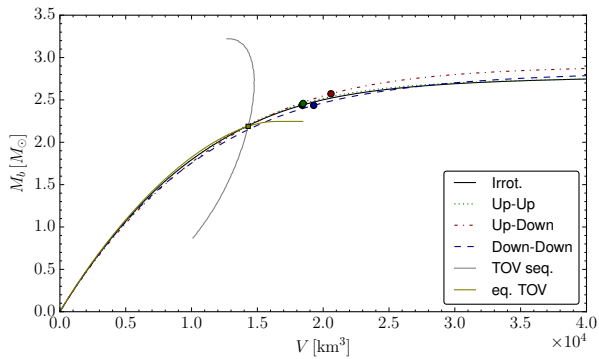

  \begin{center}
    \includegraphics[width=0.95\columnwidth]{{{mass_volume}}}  
    \caption{Baryonic mass versus proper volume contained in isodensity surfaces, 
    at $14\usk\milli\second$ after merger. The symbols mark the ``bulk'' values. 
    We also show the relation of bulk mass versus bulk volume for a sequence of
    TOV stars with the same EOS as the initial data. The intersection with the
    remnant profiles defines their ``TOV core equivalents.'' The profile of 
    the core equivalent shown in the plot corresponds to model \texttt{SHT\_IRR}.}
    \label{fig:mass_volume}
  \end{center}
\end{figure}

\begin{figure}
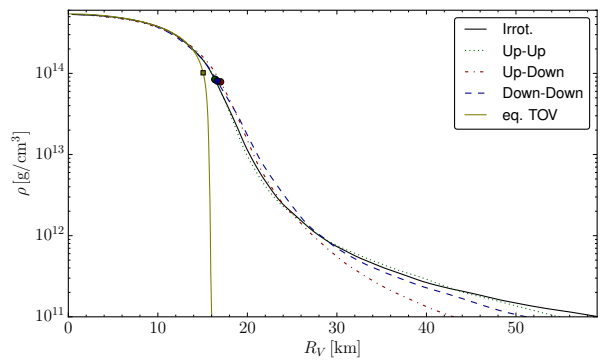

  \begin{center}
    \includegraphics[width=0.95\columnwidth]{{{dens_volume}}}  
    \caption{Density versus volumetric radius for isodensity surfaces at $14\usk\milli\second$
    after merger. The symbols mark the bulk values.
    For comparison, we also show the TOV solution approximating the 
    core (see \Fref{fig:mass_volume}).}
    \label{fig:dens_volume}
  \end{center}
\end{figure}

We now discuss the radial mass distribution of the remnants.
Figure~\ref{fig:mass_volume} depicts the mass-volume relations introduced in 
\Sref{sec:code} for all spin configurations. 
We observe differences in the core around a few percent between the different spin 
configurations, comparable to the magnitude expected from the bulk compactness shown in 
\Fref{fig:bulk_comp}.
In \cite{Kastaun:2016}, we found that the core of the remnant for irrotational model
has a mass profile very similar to that of a TOV solution. The mass-volume relation  
for this core-equivalent TOV model is shown as well in 
\Fref{fig:mass_volume}. 
In addition, \Fref{fig:dens_volume} shows the same 
data in terms of density versus the volumetric radius of the corresponding isodensity 
surface. We find that the radial mass distributions of remnant and TOV core equivalent 
agree indeed well. Centrifugal forces become important for volumetric radii 
$>14\usk\kilo\meter$ (when comparing to Figs.~\ref{fig:vortex_struct_t8} and 
\ref{fig:vortex_struct_t14} note that they use the circumferential radius instead).
This agrees with the rotation profiles which show that the matter approaches
Kepler velocity in those outer layers.

\subsection{Gravitational waves}
\label{sec:gw}

For all our simulations, we extracted the GW signal using the methods detailed in \Sref{sec:gw_extr}.
In order to reduce boundary effects when computing Fourier spectra, we first applied the 
tapering window function described in \cite{McKechan:2010:27h4020M}, with a tapering duration of 
$2\usk\milli\second$. 
The results are shown in Figs.~\ref{fig:gw_spec_jump_irr} to~\ref{fig:gw_spec_jump_apr4_em} 
(the strain is also publicly available in the supplemental material \cite{Kastaun:2017:supplemental}).
The waveforms look quite typical, in particular the short minima after merger are fairly common
for BNS mergers. Using the methods described in \Sref{sec:gw_extr}, we have shown that those
can be described as overmodulation, i.e. a zero crossing of the signal amplitude.
For all models, we detected phase jumps at the location of the first two minima,
except model \texttt{APR\_EM} where only the first minima was identified as phase jump.
The jumps are particularly visible in the phase velocity, which exhibits strong peaks 
that are not present in the jump-corrected phase.
\nocite{advLigoSens:2010,Abbott:2016:1304.0670,Virgo:noise:adv,ETSens:0810.0604,ETSens:data}

\begin{figure*}
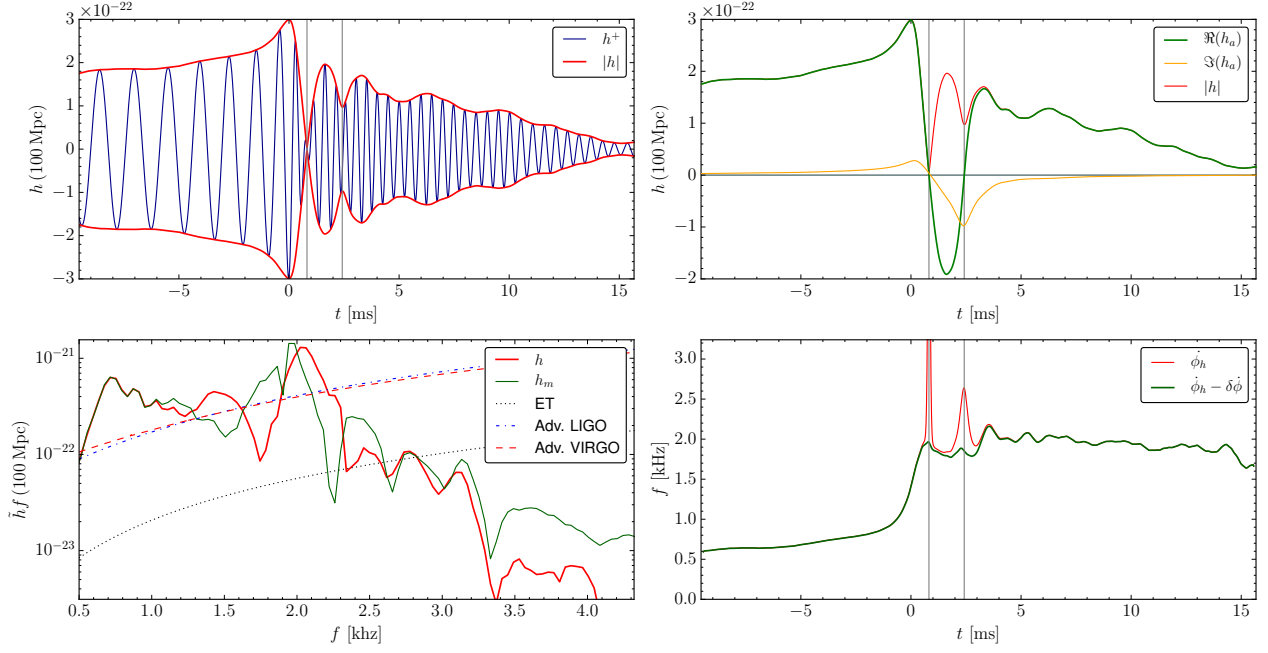

  \begin{center}
    \includegraphics[width=0.95\textwidth]{{{gw_spec_jump_irr}}}  
    \caption{GW signal for model \texttt{SHT\_IRR}.
    Upper left panel: $l=m=2$ component of the GW strain. Upper right: The 
    evolution of the complex amplitude $h_a$ defined in \Sref{sec:jump_det}, 
    indicating two phase jumps after merger (marked by the vertical lines).
    Lower right: Phase velocity both for the continuous phase $\phi_h$
    of $h$ as well as the phase $\phi_h - \delta\phi$ corrected for phase jumps.
    Lower left: Fourier spectrum of $h$ at $100\usk\mega\parsec$ compared to GW detector 
    design sensitivity
    curves [advanced LIGO, zero detuning, high power \cite{advLigoSens:2010},
    advanced Virgo \cite{Abbott:2016:1304.0670,Virgo:noise:adv}, 
    Einstein Telescope (configuration ET-B) \cite{ETSens:0810.0604,ETSens:data}]. 
    For comparison, we also show the spectrum of the strain after removing 
    the phase jump, given by $h_m = h e^{-i \delta\phi}$.}
    \label{fig:gw_spec_jump_irr}
  \end{center}
\end{figure*}

The presence of a phase jump shortly after merger makes it more difficult to interpret the 
spectra. 
In particular time-frequency maps will exhibit a decrease in amplitude near the phase jump 
since the signal before and after partially cancels. This might in principle be relevant 
for any classification scheme of GW merger events (such as
\cite{Vinciguerra:2017:094003,Clark:2014:062004}), or construction of matched filtering templates
(such as \cite{Clark:2016:085003}), if they are carried out in the frequency or time-frequency domain.
However, the improvement of GW analysis pipelines is beyond the scope of 
the current paper. Here we only try to estimate the potential impact of cancellation effects 
on the Fourier spectrum. For this, we compute also the spectrum of a modified signal given by
$h e^{-i \delta\phi}$, where $\delta\phi$ is the jump correction introduced in \Sref{sec:jump_det}.
We stress that the factor $e^{-i \delta\phi}$ affects the frequency
only in short intervals around the jumps. We therefore attribute any large changes in the spectrum to
cancellation effects between the segments separated by the jumps.

As shown in Figs.~\ref{fig:gw_spec_jump_irr} to~\ref{fig:gw_spec_jump_apr4_em}, the impact is 
dramatic. Specifically, all spectra of the standard strain show a broad peak between the frequency 
ranges of inspiral and postmerger phase, separated from the main postmerger peak by a relatively 
steep valley (e.g. at $1.75\usk\kilo\hertz$ in \Fref{fig:gw_spec_jump_irr}). The location of the
low-frequency peak and the valley change significantly. 
The effects of the phase jumps are also noteworthy since
low-frequency peaks have been interpreted as combination frequencies of $m=0$ and $m=2$ oscillations
of the remnant in \cite{Stergioulas:2011:427}, while \cite{Takami:2014:91104} correlated those peaks 
phenomenologically with the 
compactness if the initial NSs. Our findings indicate that, at least for the cases at hand, those
peaks might not be the result of one or several physical oscillations, but could be generated by the 
superposition of a phase-cancellation induced valley onto an underlying frequency distribution that is 
much smoother. Although this possibility complicates GW data analysis, it also gives reason to hope 
that the behavior of the actual frequency evolution of merger events is less complex than 
suggested by the strain, once phase jumps are accounted for. 

Now that we have established the presence of overmodulation in GW signals, we propose a model 
to explain the physical causes. For this, we consider the dynamics of the merger in a rotating frame 
with angular velocity $\Omega = \dot{\phi}/2$, where $\phi$ is the jump corrected and smoothed phase 
of the $m=2$ component of the GW strain. The exact rotation rate is not important, only that it is slowly varying and 
in phase with the GW signal on average. We then consider the $m=2$ multipole moment $Q_c$ in the 
corotating frame. The multipole moment $Q_I$ in the inertial frame is given by
$Q_I = Q_c e^{i\phi}$. We now regard $Q_c$ as a slowly varying complex-valued amplitude modulating
the oscillation given by the factor $e^{i\phi}$. Of course, $Q_c$ can undergo a sign change. For the 
$m=2$ moment, a sign change means that the principal axes are interchanged. It is plausible that this 
happens during merger. We therefore 
have all the ingredients for overmodulation of $Q_I$. The only complication is that the GW strain
is given by the second derivative of $Q_I$. For the strain $h = h_c e^{i\phi}$, we thus obtain 
\begin{align}
h_c &= \ddot{Q}_c + 2 i \omega \dot{Q}_c + \left(i \dot{\omega} - \omega^2 \right) Q_c,
\end{align}
where $\omega=\dot{\phi}$.
For the simplest case of a linear zero crossing of the form $Q_c = \dot{Q}_c \left( t - t_j \right)$,
and assuming $\dot{\omega} \approx 0$, we obtain 
$h_c = i \omega \dot{Q}_c \left(2 + i \omega\left(t-t_j\right) \right)$. This is exactly the form assumed for the 
jump fitting shown in \Sref{sec:jump_det}, with $k=\omega/2$. 
For this simple case, we find that the strain itself does not 
exhibit a zero crossing, but assumes a minimum determined by the speed of the zero crossings of $Q_c$.
However, the strain does undergo a phase shift by $\pi$. This simple model can also explain the large second
derivative of strain amplitudes often seen in gravitational waveforms, without requiring a sudden
change of the multipole moment in the rotating frame. 
We caution that some of the broader peaks in phase velocity might have different causes than near 
zero crossing of $Q_c$. For the sharp peaks, however, the frequency changes by more than $1\usk\kilo\hertz$ 
within less than $0.5\usk\milli\second$. It is implausible that the remnant's oscillation frequency or 
rotation rate changes so drastically within a period shorter than the dynamical time scale;
our model provides a more natural explanation.

For example, in \cite{Kastaun:2015:064027} we observed some strong peaks in the GW phase velocity
during merger, which we interpreted as a consequence of a compactness maximum of the remnant. After 
reanalysing the data with the new methodology, we find that the frequency peak at merger
is partially due to a phase jump. Another example is
the GW strain for model \texttt{APR4\_UM} shown in \Fref{fig:gw_spec_jump_apr4_um}, which has 
a very pronounced minimum around $3\usk\milli\second$ after merger that we left unexplained in 
\cite{Endrizzi:2016:164001}. We now find that this minimum is a very clear 
case of a phase jump, judged by the sharp peak in the instantaneous frequency. Thus, the minimum is 
caused by a simple zero crossing of the quadrupole moment. The characteristic behavior of the phase 
makes other explanations very unlikely. In particular, we rule out that the minimum is the superposition
of a damped mode excited at merger and another mode growing unstably 
or that it is a simple beating phenomenon between two modes. 
We are not aware of any GW data analysis targeting temporary minima. Nevertheless, if
postmerger signals that turn off temporarily should be observed by GW astronomy,
such considerations will become crucial for the physical interpretation.

\begin{figure*}
  \begin{center}
    \includegraphics[width=0.95\textwidth]{{{gw_spec_jump_uu}}}  
    \caption{Like \Fref{fig:gw_spec_jump_irr}, but for model \texttt{SHT\_UU}.}
    \label{fig:gw_spec_jump_uu}
  \end{center}
\end{figure*}

\begin{figure*}
  \begin{center}
    \includegraphics[width=0.95\textwidth]{{{gw_spec_jump_ud}}}  
    \caption{Like \Fref{fig:gw_spec_jump_irr}, but for model \texttt{SHT\_UD}.}
    \label{fig:gw_spec_jump_ud}
  \end{center}
\end{figure*}

\begin{figure*}
  \begin{center}
    \includegraphics[width=0.95\textwidth]{{{gw_spec_jump_dd}}}  
    \caption{Like \Fref{fig:gw_spec_jump_irr}, but for model \texttt{SHT\_DD}.}
    \label{fig:gw_spec_jump_dd}
  \end{center}
\end{figure*}

\begin{figure*}
  \begin{center}
    \includegraphics[width=0.95\textwidth]{{{gw_spec_jump_apr4_um}}}  
    \caption{Like \Fref{fig:gw_spec_jump_irr}, but for model \texttt{APR4\_UM}.}
    \label{fig:gw_spec_jump_apr4_um}
  \end{center}
\end{figure*}

\begin{figure*}
  \begin{center}
    \includegraphics[width=0.95\textwidth]{{{gw_spec_jump_apr4_em}}}  
    \caption{Like \Fref{fig:gw_spec_jump_irr}, but for model \texttt{APR4\_EM}. Note the instantaneous
    frequency variations after $30\usk\milli\second$ are artifacts of the low signal amplitude. }
    \label{fig:gw_spec_jump_apr4_em}
  \end{center}
\end{figure*}

\begin{figure}
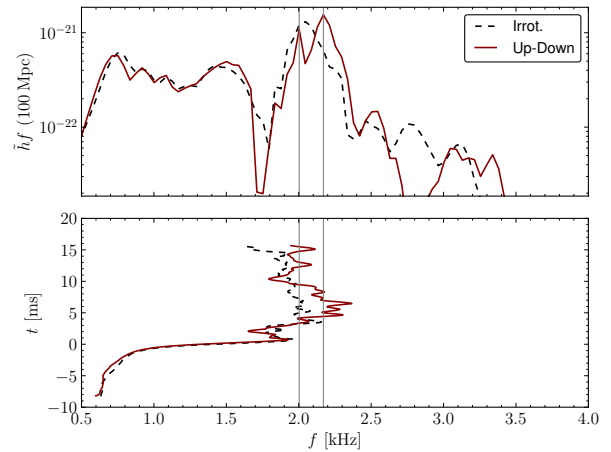

  \begin{center}
    \includegraphics[width=0.95\columnwidth]{{{gw_specs_ifreq_irr_ud}}}  
    \caption{Upper panel: GW power spectrum for irrotational and up-down spin 
    configurations. Before Fourier analysis, the signals have been cut to the 
    common time interval with respect to merger time to exclude differences 
    due to simulation length.
    Lower panel: Time evolution of the jump-corrected instantaneous 
    frequency ($t=0$ is the time of merger).
    The vertical lines mark the peaks of the power spectrum for the up-down case.}
    \label{fig:gw_specs_irr_ud}
  \end{center}
\end{figure}

\begin{figure}
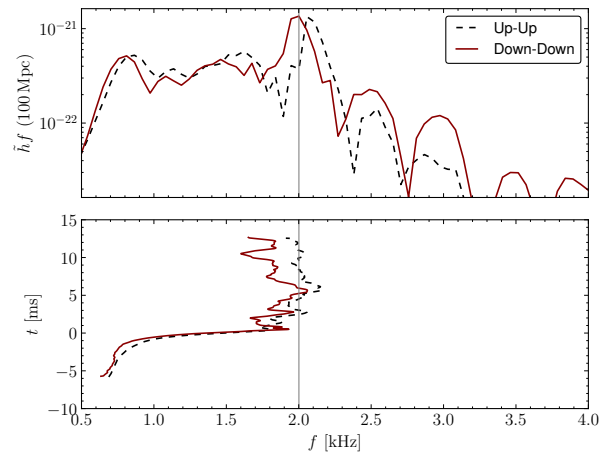

  \begin{center}
    \includegraphics[width=0.95\columnwidth]{{{gw_specs_ifreq_uu_dd}}}  
    \caption{Like \Fref{fig:gw_specs_irr_ud}, but comparing the up-up and down-down 
    spin configurations. 
    The vertical line marks the peak in the down-down spectrum.}
    \label{fig:gw_specs_uu_dd}
  \end{center}
\end{figure}

After discussing the generic GW features, we now turn to the impact of the NS spin.
Figures~\ref{fig:gw_specs_irr_ud} and~\ref{fig:gw_specs_uu_dd} show a comparison of the GW spectra
of the four cases. Further, we report in \Tref{tab:outcome} the frequency of the maximum of the 
power spectrum (excluding the part corresponding to inspiral) as well as the instantaneous 
frequency  at merger time. We estimate the finite resolution error of the main postmerger 
frequency peak around $2\%$, based on the resolution study in \cite{Ciolfi:2017:063016}.
Although the location of the main peak is affected by the spin, the shift is generally 
comparable to the width of the peak. One noteworthy feature is the appearance of a small side peak 
for model \texttt{SHT\_UD}. As seen in \Fref{fig:gw_spec_jump_ud}, this feature is not caused by 
cancellation effects related to phase jumps.
By comparing to the instantaneous frequency (the phase velocity) shown in Figs.~\ref{fig:gw_specs_irr_ud} 
and~\ref{fig:gw_specs_uu_dd}, we find that the splitting
of the main peak is caused by a small frequency change occurring around 
$8\usk\milli\second$ after merger. This coincides with a rearrangement of the vortex structure
that can be seen in Figs.~\ref{fig:vortex_struct_t8} and~\ref{fig:vortex_struct_t14}. 
One plausible explanation would be that the rearrangement of the fluid flow slightly changes the moment 
of inertia, thus causing a small decrease in the rotation rate. Note that also the fluid flows of other models 
undergo changes. The difference between the up-down and the irrotational model is that the transition happens more 
smoothly for the latter,
and correspondingly the instantaneous frequency shows a more continuous drift as well. This indicates that drifts of the 
frequency in the late postmerger phase might be caused by a slow change of the remnant's vortex distribution, offering 
an alternative explanation to the common notion that the angular momentum carried away by GWs is responsible.
We recall that angular momentum loss affects the frequency not simply by a change in rotation rate, but mostly 
by the change in compactness (see also the discussions in 
\cite{Bernuzzi:2014:104021,Dietrich:2016:044045,Dietrich:2017:024029}), 
which makes it difficult to distinguish from the effects of internal structural rearrangements.
For the gradually changing models, one could argue that the angular momentum loss is a common cause both
for changes of vortex distribution and GW frequency; the more rapid change observed for the up-down model seems to 
indicate otherwise, however.

A rearrangement of vortices might also lead to strong strain amplitude variations in some cases,
which might explain minima sometimes observed in numerical waveforms. 
In \cite{Feo:2017}, it was suggested that the rise of the GW amplitude following minima in
specific cases might be caused by the growth of unstable modes.
Evidence for such instabilities has only been found
for hypermassive neutron star (HMNS) models with rapidly rotating core however (see e.g. \cite{Shibata:2003:343,Loeffler:2015:064057}). 
The reduction in amplitude around $8$ to $10\usk \milli\second$ after merger shown in 
\Fref{fig:gw_spec_jump_ud} provides at least one example where the change can be attributed
to a change of the vortex structure that takes place simultaneously (as can be seen 
in a movie available in the supplemental material \cite{Kastaun:2017:supplemental}).
There is no reason to assume 
such a rearrangement could not produce a temporary minimum instead; in fact, 
we observed a minima coinciding with a vortex rearrangement in \cite{Ciolfi:2017:063016}
(for the unequal mass H4 model).
In this context, we also note that 
some of the waveforms presented in \cite{Hotokezaka:2013:44026} (e.g. models \texttt{ALF-120150}, 
\texttt{H4-120150}) show the sharp maxima or minima in the instantaneous frequency characteristic 
for phase jumps exactly at the location of strain minima. The relation between vortex structure and 
strain amplitude clearly deserves further study.

\subsection{Disk and matter ejection}
\label{sec:disk}
We now turn to the effect of the spin on the disk and the ejected matter.
To this end, we first compute the total mass outside a 
coordinate sphere with proper volumetric radius  
of $26\usk\kilo\meter$, defining a tentatively named ``disk mass.''
We stress that the mass depends strongly on the cutoff radius. 
Our choice of the latter was motivated 
by Fig.~16 in \cite{Kastaun:2016}, showing the isodensity contours for 
our irrotational SHT model in the meridional plane (for comparison, the
cutoff radius corresponds to $20\usk\kilo\meter$ in the simulation 
coordinates used in the figure). Further, the bulk surfaces shown in 
\Fref{fig:vortex_struct_t14} are completely contained within a coordinate 
sphere with proper volumetric radius $24\usk\kilo\meter$.

Not all of this matter really 
forms a disk. The parts at large radius typically have strongly eccentric orbits
with periods larger than the evolution time. This matter will eventually fall back 
onto the inner system and might give rise to interesting phenomena outside the scope of 
this paper. Of course, the transition between circularized disk and eccentric fallback
component is gradual. To give a number however, we computed the matter outside a 
radius of $60\usk\kilo\meter$, as in \cite{Endrizzi:2016:164001}.
The masses outside $26$ and $60\usk\kilo\meter$ are reported in \Tref{tab:outcome} 
for the SHT models. We find a considerable impact of the spin on both 
measures. 
For the models at hand, the disk mass for the up-down (down-down) case is only 
$55\%$ ($80\%$) of the mass for the up-up case (or the almost identical irrotational case). 
Although we have no convergence 
test, we are optimistic that those differences are larger than the numerical error.
The inner parts of the disk are also visible in \Fref{fig:dens_volume}, 
showing density versus volumetric radius of the corresponding isodensity surface.
Here we find that the density falls off most strongly for model \texttt{SHT\_UD}.

In this context it is worth noting that the matter in the disk is not 
formed completely during merger, but material is migrating outwards continuously. 
In \cite{Kastaun:2016} we have shown for the irrotational model that many fluid 
trajectories ending up in the disk interact with secondary vortices at some point.
Since the latter are sensitive to the spin, the impact of the spin on the disk mass 
is not necessarily caused only by the different centrifugal forces during the merger.
Lacking a clear measure, we did however not distinguish the amount of matter migrating 
to the disk immediately from the amount thrown out later.

In order to estimate the amount of ejected matter, we use the same tools as in
\cite{Endrizzi:2016:164001} (see also the discussion in \cite{Kastaun:2015:064027}). 
In short, we compute the time integrated 
flux of unbound matter according to the geodesic criterion through several spherical 
surfaces. As explained in \cite{Endrizzi:2016:164001}, we use the maximum obtained 
from the different surfaces ($r=111$, $148$, $222$, $295$, $443$, $591$, 
$916 \usk\kilo\meter$ for the SHT models) as a best estimate. The results are given in \Tref{tab:outcome}.
For a more direct comparison of results for the different spins, we also report
the ejected mass measured using the same extraction sphere at fixed radius $r=222\usk\kilo\meter$.
We find that all SHT models
eject only small amounts of matter ${\lesssim}10^{-3} M_\odot$. There are however strong
differences between the different spin configurations.
Most matter is ejected for the up-up case, and least for the irrotational and the 
mixed cases. We note that this is opposite to the relation found in \cite{Kastaun:2015:064027},
for a model with the SHT EOS, but in the lower hypermassive mass range. In this case,
the irrotational model ejected more matter than the up-up model. This was connected
to the ejection mechanism: the strong radial oscillations of the remnant launched 
many waves that kicked matter out of the disk, and the strength of the oscillations 
was greater for the irrotational case.

Note the accuracy of the ejected mass is difficult to estimate.
There are several sources of error: the validity of the geodesic assumption, 
the finite resolution, and interaction with the artificial atmosphere.
For the resolution error, we again use the resolution studies for magnetized
versions of the APR4 models in \cite{Ciolfi:2017:063016,Endrizzi:2016:164001},
which provided estimates around $17\%$ and $50\%$ for the equal and unequal mass
models, respectively. Lacking a dedicated convergence test, we assume 
a finite resolution error around $50\%$ for the SHT models as well. 
Regarding the artificial atmosphere, we should note that the total baryon mass 
of the system increases by up to $10^{-3} M_b^\text{tot}$ (i.e. more than the 
ejected mass) during the SHT EOS simulations. 
However, this number refers to the whole computational domain and cannot be 
added to the error of the ejecta mass. The interaction of ejected matter 
with the artificial atmosphere actually decreases the unbound mass. This was found 
by monitoring the radial distribution of ejected matter using histograms and is 
also indicated by an animation for model \texttt{SHT\_UU} showing that the matter 
becomes bound again at a large radius while expanding into the artificial atmosphere.
Using the same animation (available in the supplemental material \cite{Kastaun:2017:supplemental}) we also verified 
that the unbound matter traverses refinement 
boundaries without visible artifacts. 
All in all, we roughly estimate the mass to be accurate within a factor of 2, 
unless the ejecta mass is below $10^{-4} \usk M_\odot$, in which case we only provide 
an upper limit.

To identify the ejection mechanism for the up-up model of the present work,
we produced movies visualizing unbound matter together 
with density or temperature  in the meridional and equatorial planes. 
We found that, for this model, almost all matter is ejected in a single
wave consisting of two concentric rings above and below the orbital plane. Those
animations also gave the impression that the matter was not tidally ejected.
Instead, it was ejected from the remnant during merger, but still orbiting it,
when a shock wave originating from the remnant oscillations finally liberated the 
material. We validated this picture by tracking ejecta trajectories backward in time,
in the same fashion described for the matter trajectories in the orbital plane,
but using 3D data. Using this method, we also found that the average ejecta temperature
was increased from ${\approx}1\usk\mega\mathrm{eV}$ to ${\approx}2.5\usk\mega\mathrm{eV}$
when receiving the final kick that liberated the matter. The specific entropy was increased
as well, which implies shock heating. Subsequently, the ejecta cooled down again adiabatically.
The implications for nucleosynthesis will be discussed in a future publication;
here we just recall that the initial temperature might be important for the final abundances 
of heavy elements produced by the r-process nucleosynthesis. At the time the ejected material became 
classified as unbound according to the geodesic criterion, the temperature was already 
decreased by adiabatic expansion. It therefore seems prudent to track ejected material
back in time to get the full thermal history.

\section{Summary and conclusions}
\label{sec:summary}

We evolved the merger of binaries consisting of two $1.4\usk M_\odot$ NSs with different combinations 
of aligned and antialigned initial NS spins, as well as the irrotational case. 
Our models employ the Shen-Horowitz-Teige EOS, for which the 
merger at the given total mass results in a stable neutron star.
We considered moderate rotation rates of $164\usk\hertz$. If such rates 
are reached in nature for merging binaries is an open question; here we studied the possible 
consequences if this is the case. In particular, we investigated the inspiral time, the fluid flow in 
the remnant, the GW signal, and the mass ejection.

The inspiral time clearly depends on the spin. If both stars are aligned, the inspiral takes around
two orbits longer than for a system where both are antialigned. The irrotational and mixed alignment
cases are in between, the latter merging slightly quicker. We caution that our initial
data has some residual eccentricity. However, the same trend has been observed in several other studies
\cite{Tsatsin:2013:64060,Kastaun:2015:064027,Bernuzzi:2014:104021,Dietrich:2016:044045} which use completely 
different methods of constructing initial data with spin. In particular, the mixed system studied in
\cite{Dietrich:2016:044045} merged faster than the irrotational one, as in our case. As pointed out by
\cite{Dietrich:2016:044045}, the post-Newtonian spin-orbit coupling should cancel for the mixed 
case (with equal absolute spin), which means the difference is due to terms accounting for the 
spin-spin coupling and spin self-coupling. Note however there is also the possibility that tidal 
effects are modified by the spin, as discussed in \cite{Pani:2015:124003}.

Regarding the remnant evolution, we find the same general picture as in \cite{Kastaun:2016}: in a coordinate frame corotating
with the main $m=2$ density perturbation, we observe a slowly evolving pattern exhibiting a strong nonlinear
density perturbation, and a fluid flow consisting initially of two large vortices which slowly merge into one
vortex that could be described as strongly deformed differential rotation. However, the fluid flow also 
features  secondary vortices that are phase locked with the main density deformation and coincide with the 
location of hot spots. The secondary vortices in our simulations persist at least $20\usk\milli\second$; 
note however that we did not include magnetic fields, which, depending on the strength, might reduce the lifetime.

The spin strongly influences the size, shape and distribution of the vortices. In particular, the mixed
spin case resulted in a very asymmetric pattern. The structures are not completely stationary, but change 
gradually towards a more axisymmetric state. For the mixed case, this change took place more rapidly, 
causing also a small shift of the GW frequency around $8\usk\milli\second$ after merger.
This case is also a hint that the rearrangement is not merely a reaction to the gradual angular momentum loss
due to GWs, and that the remnant might possess a complex internal dynamics, at least in the early postmerger
phase relevant to GW astronomy.
We also analyzed the fluid flow of an unequal mass system employing the APR4 EOS, and unsurprisingly found an
asymmetric vortex structure. Finally, we performed a long-term evolution of an equal mass binary with the APR4 EOS
and a similar total mass, and found that $40\usk\milli\second$ after merger all secondary vortices had decayed,
with a fluid flow described by differential rotation. This stage is however not relevant to GW astronomy since 
the strain amplitude is already too small.

The overall rotation profile in the equatorial plane for all our models showed a maximum in the outer layers
of the remnant, while the core rotated slowly and the matter further out gradually approached Kepler 
velocity. We also found that the maximum rotation rate is given by the rotation rate of the density 
deformation, which in turn determines the main GW frequency. This behavior was already found for many different 
models in \cite{Kastaun:2015:064027, Endrizzi:2016:164001, Kastaun:2016, Hanauske:2016} and seems to hold
regardless of mass, mass ratio, EOS, and spin. 
It becomes increasingly clear that the cores of merger remnants typically do not rotate fast enough to have
a significant impact on the radial structure, and that the stability and lifetime are determined by the 
evolution of the outer layers and possibly the disk.
In this work, we measured the radial mass distribution based on a new measure introduced in \cite{Kastaun:2016}
and find indeed that the core can be approximated very well by the core of a TOV solution.

The main effect of the spin on the GW signal emitted by our models was the different length of the inspiral phase.
The frequency changes in the postmerger phase were smaller than the width of the main peak in the Fourier spectrum.
For the mixed model, we noticed the appearance of a side peak that was apparently caused by the aforementioned 
rearrangement of the fluid flow. 

Unrelated to our study of the spin, we also investigated a generic feature of 
BNS GW waveforms, namely the presence of sudden phase jumps during and/or after merger. We explain
those in terms of overmodulation, i.e. we regard the GW as an amplitude-modulated signal where the modulation
amplitude can have zero crossings. In the context of gravitational waves, the amplitude modulation is given by
the remnant's quadrupole moment in a rotating frame, and the carrier frequency to the rotation rate of 
said frame, which is chosen such that the respective quadrupole moment is varying slowly.
In all cases, we observed a phase jump during merger. This corresponds to an exchange of the principal 
axes of the mass distribution in the corotating frame when the stars collide, which is very plausible. 
We have demonstrated that cancellation effects due to the phase jumps can have a strong impact
on the GW power spectra, at least with regard to secondary peaks, and should therefore be considered 
for any GW analysis in frequency space. Gradual near-zero crossings of the quadrupole moment in the 
rotating frame might also explain some of the minima in the strain often found in numerical studies. 
For one such case we could rule out other explanations for the rebrightening of the GWs, such as 
unstable mode growth, since the amplitude minima was accompanied by a clear phase jump.

Finally, we studied the mass ejection and found a strong influence of the spin. The configuration
with both spins aligned ejected the largest amount of matter, around $10^{-3} M_\odot$. Note that
the influence of spin depends also on the ejection mechanism. In our case, we found a single wave of 
matter consisting of material ejected from the disk by a shock wave originating from the remnant.
The matter was heated by the shock to around $2\usk\mega\electronvolt$ and subsequently cooled 
by adiabatic expansion. In \cite{Kastaun:2015:064027}, we studied heavier models with the same EOS 
and found that the irrotational model ejected more matter than the aligned one. In this case, matter 
was ejected in several waves
caused by the strong radial pulsations of the remnant, which in turn were stronger in the irrotational case.
For the unequal mass systems studied in \cite{Dietrich:2016:044045}, the matter was ejected tidally and
the amount depended on the spin of the lighter star. Comparing those studies, we conclude that
the initial spin has a strong influence, but whether a given spin leads to more or less ejecta is hard
to predict due to the different ejection mechanisms.

\acknowledgments
We acknowledge support from MIUR FIR Grant No.~RBFR13QJYF.
Numerical calculations have been made possible through a CINECA ISCRA 
class B Grant providing access to the FERMI cluster, and a CINECA-INFN
agreement, providing access to resources on GALILEO and MARCONI at CINECA. 
We acknowledge PRACE for awarding us access to resource SUPERMUC based in 
Germany at LRZ (Grant GRSimStar). We also like to thank the referee 
for many valuable suggestions.

\bibliographystyle{apsrev4-1-noeprint}
\bibliography{trento}

\end{document}